\documentclass[12pt,onecolumn,oneside,a4paper,peerreview]{IEEEtran}
\usepackage{cite}
\usepackage{graphicx}
\usepackage{amsmath}
\usepackage{times}
\usepackage{latexsym}
\usepackage{graphicx}
\usepackage{bm}
\usepackage{amssymb}
\usepackage[center]{caption2}
\usepackage{stfloats}
\usepackage{cases}
\usepackage{subfigure}
\usepackage{array}
\usepackage{setspace}
\usepackage{fancyhdr}
\usepackage{cite}
\usepackage{color}

\newtheorem{theorem}{Theorem}

\newtheorem{corollary}{Corollary}
\newtheorem{proposition}{Proposition}

\def\proof{\noindent\hspace{2em}{\itshape Proof: }}
\def\endproof{\hspace*{\fill}~$\square$\par\endtrivlist\unskip}

\newcaptionstyle{mystyle2}{%
\captionlabel $.$ \, \doublespacing \captiontext \par}
\captionstyle{mystyle2}

\begin{document}
\title{Optimization and Analysis of Wireless Powered Multi-antenna Cooperative Systems}
\author{{Han Liang, \emph{Student Member}, \emph{IEEE},
                      Caijun Zhong, \emph{Senior Member}, \emph{IEEE},
                      Himal A. Suraweera, \emph{Senior Member}, \emph{IEEE},
                      Gan Zheng, \emph{Senior Member}, \emph{IEEE},
                      and Zhaoyang Zhang}, \emph{Member}, \emph{IEEE}

\thanks{H. Liang, C. Zhong and Z. Zhang are with the Institute of Information and Communication Engineering, Zhejiang University, China (email: caijunzhong@zju.edu.cn).}
\thanks{H. A. Suraweera is with the Department of Electrical \& Electronic Engineering, University of Peradeniya, Peradeniya 20400, Sri Lanka (e-mail: himal@ee.pdn.ac.lk).}
\thanks{G. Zheng is with  the Wolfson School of Mechanical, Electrical and Manufacturing Engineering, Loughborough University, Leicestershire, LE11 3TU, UK. (e-mail: g.zheng@lboro.ac.uk).}}

\maketitle
\begin{abstract}
In this paper, we consider a three-node cooperative wireless powered communication system consisting of a multi-antenna hybrid access point (H-AP) and a single-antenna relay and a single-antenna user. The energy constrained relay and user first harvest energy in the downlink and then the relay assists the user using the harvested power for information transmission in the uplink. The optimal energy beamforming vector and the time split between harvest and cooperation are investigated. To reduce the computational complexity, suboptimal designs are also studied, where closed-form expressions are derived for the energy beamforming vector and the time split. For comparison purposes, we also present a detailed performance analysis in terms of the achievable outage probability and the average throughput of an intuitive energy beamforming scheme, where the H-AP directs all the energy towards the user. The findings of the paper suggest that implementing multiple antennas at the H-AP can significantly improve the system performance, and the closed-form suboptimal energy beamforming vector and time split yields near optimal performance. Also, for the intuitive beamforming scheme, a diversity order of $\frac{N+1}{2}$ can be achieved, where $N$ is the number of antennas at the H-AP.
\end{abstract}

\newpage
\section{Introduction}\label{section:introduction}
The advancement of rectenna technology in recent years has significantly boosted the efficiency of far-field wireless power transfer (WPT), thereby paving the way for its practical deployment in various applications \cite{M.Xie}. One particularly promising scenario is wireless communications networks, where the prospect of cutting the last wire and enabling perpetual operation time for wireless devices by WPT has attracted significant interests from both industry and academia.

The integration of WPT into wireless communications networks has resulted in a newly emerged research frontier, namely, wireless powered communication network (WPCN), where energy constrained wireless devices are first powered by WPT and then perform information transmission using the harvested radio frequency (RF) energy \cite{S.Bi}.\footnote{Another paradigm is simultaneous wireless information and power transfer (SWIPT), where the H-AP communicates with the active nodes in the cell, while the remaining idle nodes scavenge energy from the transmitted RF signals by the H-AP \cite{R.Zhang0}.} In the literature, two promising WPCN architectures have been widely adopted, i.e., hybrid access point (H-AP) architecture, where the AP not only participates in the information transmission, but also acts as an energy source to power the mobile devices \cite{K.Huang2,H.Ju0}, and the power beacon architecture, where dedicated power beacons are deployed in the network to power mobile devices \cite{K.Huang}.

While WPCN enjoys great flexibility enabled by WPT, its performance is critically depended on the energy harvested at the nodes. Thus an important question is what is the fundamental information theoretical limitation of WPCN? To answer this question, so far the performance of WPCN has been investigated in different settings and protocols. In \cite{H.Ju}, a harvest-then-transmit (HTT) protocol was proposed, where the users first collect energy from the signals broadcasted by the H-AP in the downlink, and then use the harvested energy to transmit independent information to the H-AP in the uplink according to the time-division-multiple-access (TDMA) mechanism. Based on the HTT protocol, \cite{Y.L.Che} investigated the performance of a large-scale WPCN with wireless nodes equipped with/without battery. Optimal resource allocation for WPCN was studied in \cite{Q.Wu}, while in \cite{X.Chen0,L.Liu2,C.Zhong,W.Huang,C.Zhong0}, the throughput performance of WPCN employing energy beamforming was investigated.

To further improve the performance of WPCNs, the idea of cooperative transmission was proposed in \cite{H.Ju3}, where users first harvest energy from the H-AP in the downlink energy transmission phase, and then work cooperatively during the uplink information transmission phase. Recently, for the elementary three-node WPCN, a harvest-then-cooperate (HTC) protocol was proposed in \cite{H.Chen}, where the energy constrained source and relay harvest energy in the downlink, and then the relay assists the source's information transmission in the uplink. It was shown that the proposed HTC protocol significantly improves the average throughput of the system compared to a point-to-point system with no relay. One common limitation of the prior works on cooperative WPCNs is that they all assume single antenna H-AP. Since energy efficiency is a critical concern for WPCNs, and implementing multiple antenna can help to substantially improve the efficiency of WPT, in addition to increase the communication capacity by exploiting the spatial degree of freedom, it is expected that multiple antenna techniques will be an indispensable component for future WPCNs \cite{Z.Ding}. However, empowering H-AP with multi-antennas also brings fundamental challenges for the design and analysis of WPCNs, i.e., how to design optimal energy beamforming vectors such that a fine energy balance can be maintained between the user and relay, and how to characterize the achievable system performance.

To answer the above questions, we extend the model considered in \cite{H.Chen} by equipping multiple antennas at the H-AP. In addition, to further improve the system performance, instead of the selection combining (SC) technique adopted in \cite{H.Chen}, the maximal ratio combining (MRC) scheme is used. The optimal energy beamforming vector ${\bf w}$ and time split $\tau$ maximizing the achievable throughput is first studied.\footnote{Part of the results have been presented in the conference paper \cite{SigTel}. However, the detailed proof of the key results is not included due to space limitation, which will be presented here. On top of the suboptimal design studied in \cite{SigTel}, the optimal design and asymptotic large antenna design will be conducted. Moreover, a detailed performance in terms of throughput and outage probability has been presented in the current paper.} Due to the non-convex nature of the optimization problem, we propose a sequential optimization approach, and devised a relatively simple solution for the optimal ${\bf w}$ and $\tau$ requiring only a two-dimensional search. To further reduce the computational complexity, we propose a simple upper bound for the achievable throughput, and then obtain closed-form expressions for the optimal ${\bf w}$ and $\tau$ maximizing the throughput upper bound. In addition, in the asymptotic large antenna regime, a more compact solution for the optimal ${\bf w}$ is derived by exploiting the law of large numbers. Finally, for the special scenario where the H-AP directs all the energy towards the user, we present a detailed performance analysis on the outage probability and average throughput of the system.

The rest of paper is organized as follows: Section \ref{system model} introduces the system model. Section \ref{section:optimal beamformer} addresses the optimal beamformer and time split design problem. Section \ref{section:suboptimal beamformer} considers the suboptimal beamformer and time split design. Section \ref{section:performance analysis} investigates the outage probability and the average throughput. Numerical results are presented in Section \ref{numerical result}. Finally, Section \ref{section:conclusion} concludes the paper and summarizes the main findings.

{\em Notation:} The upper bold case letters, lower bold case letters and lower case letters denote matrices, vectors and scalars, respectively. For a complex vector ${\bf{h}}$, ${\bf{h}}^T$, ${\bf{h}}^*$ and ${\bf{h}}^{H}$ denote the transpose, conjugate and conjugate transpose of ${\bf{h}}$ respectively. $\left\|{{\bf{h}}}\right\|$ denotes the Frobenius norm of a complex vector ${\bf{h}}$. ${\bf{I}}$ denotes the identity matrix, and $\Pi_{{\bf{X}}}={\bf{X}}\left({\bf{X}}^{H}{\bf{X}}\right)^{-1}{\bf{X}}^{H}$ denotes the orthogonal projection onto the column space of ${\bf{X}}$, and $\Pi_{{\bf{X}}}^{\bot}={\bf{I}}-\Pi_{{\bf{X}}}$ denotes the orthogonal projection onto the orthogonal complement of column space ${\bf{X}}$. ${F}_t(x)$ denotes the cumulative distribution function (CDF) of a random variable $t$. $\mathbb{E} \left\{ \cdot \right\}$ denotes the statistical expectation. ${{\cal CN} (0,1)}$ denotes a scalar complex Gaussian distribution with zero mean and unit variance. $\Gamma(x)$ is the gamma function\cite[Eq. (8.310.1)]{Tables}. $\psi(x)$ is the psi function\cite[Eq (8.360)]{Tables}. $K_n\left(x\right)$ is the \emph{n}-th order modified Bessel function of the second kind \cite[Eq. (8.407)]{Tables}.
$G_{p,q}^{m,n}(\cdot)$ is the Meijer's \emph{G}-function\cite[Eq. (9.301)]{Tables}.

\section{system model}\label{system model}

We consider a three-node wireless powered communication system where a H-AP with $N$ antennas communicates with a single-antenna user with the assistance of a single-antenna relay as illustrated in Fig. 1(a). It is assumed that the relay and user have no external power supplies and both are wireless charged by harvesting energy from the H-AP. We consider the block Rayleigh fading channel model, hence the channels remain constant during one transmission block and vary independently from one block to the other. \footnote{Considering the relatively short distances between the nodes in WPT systems, Rician fading model is an appropriate choice. On the other hand, according to \cite{C.Zhong}, the line-of-sight effect is rather insignificant, especially in a multiple antenna setting. Therefore, as in many prior works such as \cite{H.Ju}, we assume Rayleigh fading model in the current work.} Also, we assume the H-AP has perfect channel state information (CSI) of the system as in \cite{Y.Zeng2}, the relay has perfect CSI of the user to the relay channel and the relay to the H-AP channel.\footnote{In practice, the CSI can be acquired through various methods, e.g., pilot-assisted reverse-link channel training \cite{Y.Zeng,F.Gao}. Nevertheless, it is also of interest to study the design and analysis of cooperative wireless powered communications taking into account of imperfect CSI.}

\begin{figure}[ht]
  \centering
  \subfigure[]{\label{fig:1a}\includegraphics[scale=0.4]{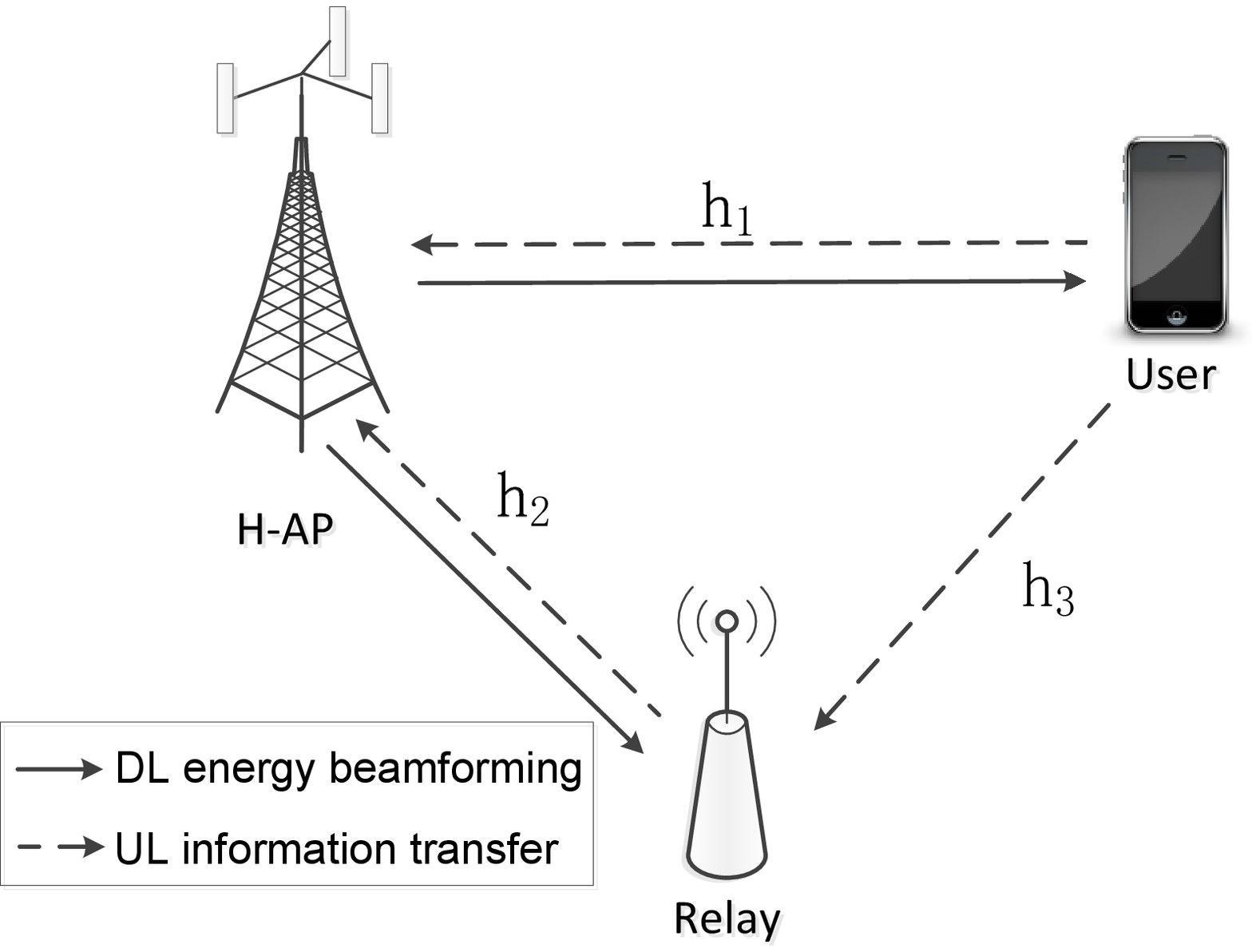}}
  \subfigure[]{\label{fig:1b}\includegraphics[scale=0.4]{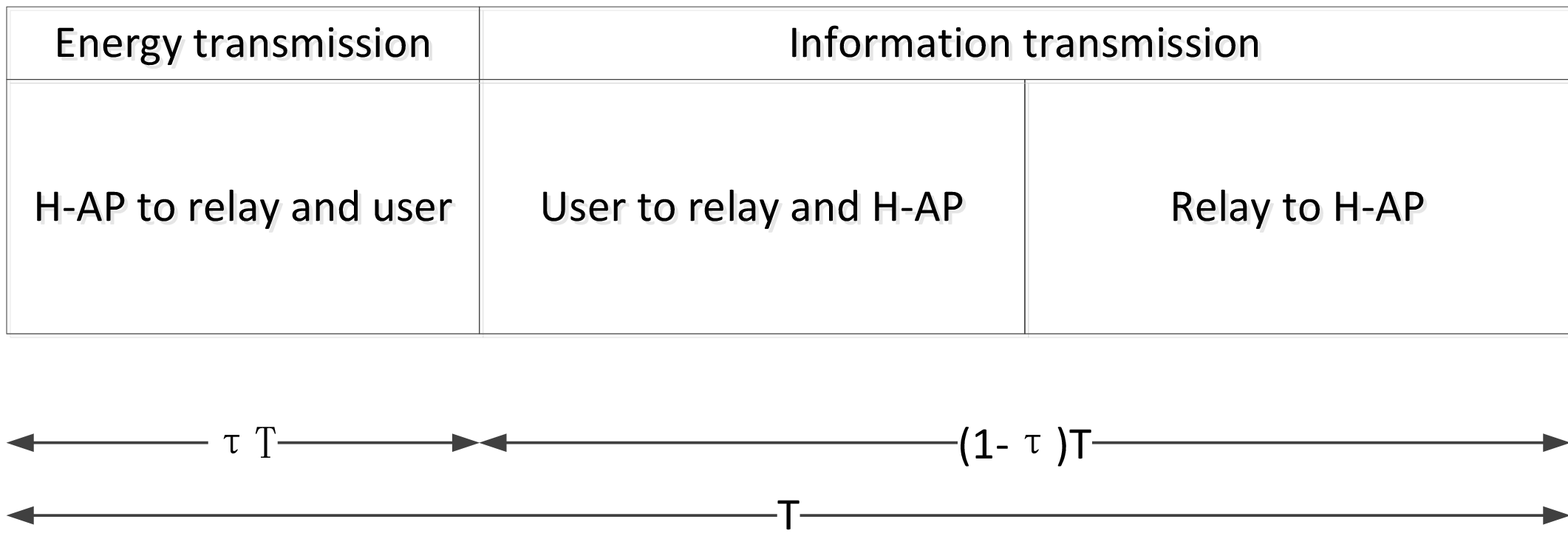}}
  \caption{(a) System model.   (b) HTC transmission protocol.}
  \label{fig:1}
\end{figure}

The relay operates in the half-duplex (HD) mode. Also, to reduce the implementation cost and energy consumption, the AF protocol is used. For the considered HTC protocol shown in \ref{fig:1b}, the entire transmission block $T$ is divided into three separate phases, namely, energy harvesting phase of length \(\tau T\) with $0<\tau<1$ being the time split parameter, where the H-AP transmits energy signals to power both the user and the relay; information transmission phase with duration \(\frac{(1-\tau) T}{2}\), where the user broadcasts the information to the H-AP and the relay; and relaying phase with duration \(\frac{(1-\tau) T}{2}\), where the relay helps to forward the received information to the H-AP.

At the end of the first phase, the total harvested energy by the user can be expressed as \cite{A.Nasir}\footnote{In this paper, we assume that the received energy is sufficient to activate the energy harvesting circuit as in \cite{H.Chen,A.Nasir}.}
\begin{align}
E_u=\frac{\eta_1\tau TP_{S}\left|{\bf{h}}_1^T {\bf{w}}\right|^2}{d_1^{\alpha_1}},
\end{align}
where ${\bf{w}}$ is the energy beamforming vector used by the H-AP, the $N\times 1$ vector ${\bf{h}}_1$ represents the user to H-AP channel with entries being independent and identically distributed (i.i.d.) ${{\cal CN} (0,1)}$ random variables, and $\eta_1$ denotes the RF energy conversion efficiency at the user, while $P_{S}$ is the transmit power of the H-AP and $d_1$ is the distance from the H-AP to the user with $\alpha_1$ being the path loss exponent.

Similarly, the total harvested energy by the relay can be expressed as
\begin{align}
E_r=\frac{\eta_2\tau TP_{S}\left|{\bf{h}}_2^T {\bf{w}}\right|^2}{d_2^{\alpha_2}},
\end{align}
where the $N\times 1$ vector ${\bf{h}}_2$ denotes the the relay to the H-AP channel, with entries being i.i.d. ${{\cal CN} (0,1)}$ random variables, and $\eta_2$ denotes the RF energy conversion efficiency at the relay, and $d_2$ is the distance from the H-AP to the relay with $\alpha_2$ being the path loss exponent. Without loss of generality, in the sequel, we assume $\eta_1=\eta_2=\eta$, and $\alpha_1=\alpha_2=\alpha$.


As in many prior works such as \cite{A.Nasir,H.Chen}, we assume that the harvested energy is fully used for information transmission.\footnote{Note that, as in \cite{H.Chen,Y.Huang1}, the power required for CSI acquisition and circuit operation is not supplied by the harvested energy, and may come from an independent battery.} Hence, the transmit power of the user and the relay can be expressed as
\begin{align}\label{puser}
P_{u}=\frac{E_{u}}{\frac{1-\tau}{2}T}=\frac{2\eta\tau P_s\left|{\bf{h}}_1^T {\bf{w}}\right|^2}{(1-\tau)d_1^\alpha},
\end{align}
and
\begin{align}\label{prelay}
P_{r}=\frac{E_{u}}{\frac{1-\tau}{2}T}=\frac{2\eta\tau P_s\left|{\bf{h}}_2^T {\bf{w}}\right|^2}{(1-\tau)d_2^\alpha},
\end{align}
respectively.

In the second phase, the user broadcasts the information to the relay and the H-AP. {Assuming channel reciprocity,} the signal received by the relay and the H-AP can be written as
\begin{align}
y_{ur}&=\sqrt{\frac{P_{u}}{d_3^\alpha}}h_3x+n_r,
\end{align}
and
\begin{align}\label{snr2:1}
{\bf{y}}_{us}&=\sqrt{\frac{P_{u}}{d_1^\alpha}}{\bf{h}}_1x+{\bf{n}}_{s},
\end{align}
respectively, where $h_3$ denotes the user-to-relay channel, $d_3$ is the distance between the user and the relay, $x$ is the information symbol with $\mathbb{E}\{xx^{*}\}=1$, $n_r$ is the additive white Gaussian noise (AWGN) at the relay with ${\mathbb{E}\{n_r n_r^{*}\}} = N_0$, while ${\bf{n}}_{s}$ is the AWGN at the H-AP with ${\mathbb{E}\{{\bf{n}}_{s}{\bf{n}}_{s}^{H}\}} = N_0{\bf{I}}$.

Finally, in the third phase, the relay amplifies and forwards the received signal to the H-AP, and the transmitted signal at the relay can be expressed as
\begin{align}
{x}_{ur}=\beta y_{ur},
\end{align}
where $\beta$ is a scaling factor to meet the power constraint at the relay, and is given by
\begin{align}
\beta^2=\frac{P_r}{\frac{P_u}{d_3^\alpha}\left|h_3\right|^2+N_r}.
\end{align}
As such, the received signal at the H-AP can be written as
\begin{align}\label{snr2:2}
{\bf{y}}_{rs}=\frac{1}{\sqrt{d_2^\alpha}}{\bf{h}}_2 x_{ur}+{\bf{n}}_s.
\end{align}

To detect the user symbol, the H-AP applies the MRC principle to combine the received signals according to (\ref{snr2:1}) and (\ref{snr2:2}). Hence, the effective end-to-end signal to noise ratio (SNR) at the H-AP can be formulated as
\begin{align}\label{total SNR}
\gamma_{\sf s}=\frac{2\eta\tau \rho \left|{\bf{h}}_1^T {\bf{w}}\right|^2\left\|{\bf{h}}_1\right\|^2}{(1-\tau)d_1^{2\alpha}}
+\frac{\frac{4\eta^2\tau^2\rho^2\left|{\bf{h}}_1^T {\bf{w}}\right|^2\left|{\bf{h}}_2^T {\bf{w}}\right|^2\left|h_3\right|^2\left\|{\bf{h}}_2\right\|^2}{(1-\tau)^2d_1^\alpha d_2^{2\alpha}d_3^\alpha}}{\frac{2\eta\tau \rho\left|{\bf{h}}_1^T {\bf{w}}\right|^2\left|h_3\right|^2}{(1-\tau)d_1^\alpha d_3^\alpha }+\frac{2\eta\tau \rho\left|{\bf{h}}_2^T {\bf{w}}\right|^2\left\|{\bf{h}}_2\right\|^2}{(1-\tau)d_2^{2\alpha}}+1},
\end{align}
where $\rho=\frac{P_s}{N_0}$. Therefore, the achievable throughput can be written as
\begin{align}\label{R}
R=\frac{1-\tau}{2}\log_2\left(1+\frac{2\eta\tau \rho\left|{\bf{h}}_1^T {\bf{w}}\right|^2\left\|{\bf{h}}_1\right\|^2}{(1-\tau)d_1^{2\alpha}}+
\frac{\frac{4\eta^2\tau^2\rho^2\left|{\bf{h}}_1^T {\bf{w}}\right|^2\left|{\bf{h}}_2^T {\bf{w}}\right|^2\left|h_3\right|^2\left\|{\bf{h}}_2\right\|^2}{(1-\tau)^2d_1^\alpha d_2^{2\alpha}d_3^\alpha}}{\frac{2\eta\tau \rho\left|{\bf{h}}_1^T {\bf{w}}\right|^2\left|h_3\right|^2}{(1-\tau)d_1^\alpha d_3^\alpha}+\frac{2\eta\tau \rho\left|{\bf{h}}_2^T {\bf{w}}\right|^2\left\|{\bf{h}}_2\right\|^2}{(1-\tau)d_2^{2\alpha}}+1}\right),
\end{align}
which indicates that the throughput performance relies heavily on the choice of $\tau$ and ${\bf w}$.

\section{The optimal energy beamforming vector design}\label{section:optimal beamformer}
In this section, we consider the design of the optimal energy beamforming vector ${\bf w}$ and time split parameter $\tau$ to maximize the achievable throughput of the system. It is important to note that the choice of ${\bf w}$ affects both energy harvested at the relay and user, hence the design of ${\bf w}$ is not a trivial task. Specifically, the optimization problem can be formulated as
\begin{align}
P0:&\underset{{\bf{w}}, \tau}\max~~~R\label{eqn:rate0}\\
&\mathrm{s.t.} \quad \left\|{\bf{w}}\right\|^2=1, \tau \in [0,1].
\end{align}

The optimization problem $P0$ is a non-convex problem, and the optimal design requires the joint optimization of ${\bf{w}}$ and $\tau$, which is in general difficult. To simplify this problem, we adopt a sequential optimization approach, namely, for a fixed $\tau$, we optimize ${\bf w}$. In this case, maximizing the achievable throughput becomes equivalent to the maximization of the end-to-end SNR. Hence, the optimal ${\bf w}$ becomes the solution of the following optimization problem
\begin{align}
P1:&\underset{{\bf{w}}}\max\left\{A_0\left|{\bf{h}}_1^T {\bf{w}}\right|^2+\frac{B_0\left|{\bf{h}}_1^T {\bf{w}}\right|^2\left|{\bf{h}}_2^T {\bf{w}}\right|^2}{C_0\left|{\bf{h}}_1^T {\bf{w}}\right|^2+D_0\left|{\bf{h}}_2^T {\bf{w}}\right|^2+1}\right\}\label{eqn:snr1}\\
&\mathrm{s.t.} \quad \left\|{\bf{w}}\right\|^2=1,
\end{align}
where $A_0=\frac{2\eta\tau\rho\left\|{\bf{h}}_1\right\|^2}{(1-\tau)d_1^{2\alpha}}$, $B_0=\frac{4\eta^2\tau^2\rho^2\left|h_3\right|^2\left\|{\bf{h}}_2\right\|^2}{(1-\tau)^2d_1^\alpha d_2^{2\alpha}d_3^\alpha}$, $C_0=\frac{2\eta\tau\rho\left|h_3\right|^2}{(1-\tau)d_1^\alpha d_3^\alpha}$ and $D_0=\frac{2\eta\tau\rho\left\|{\bf{h}}_2\right\|^2}{(1-\tau)d_2^{2\alpha}}$.

The optimization problem $P1$ is still a non-convex problem, which requires an $N$-dimensional search, hence is difficult to solve. However, it turns out that $P1$ can be converted to an equivalent problem requiring a simple one-dimensional search. To show this, we first present the following proposition.

\begin{proposition}
The optimal beamformer vector $\bf{w}$ for the optimization problem $P1$ has the following structure
\begin{align}\label{optw2}
{\bf{w}}_{opt}=\bar{x}\frac{\Pi_{{\bf{h}}_1^{*}}{\bf{h}}_2^{*}}{\left\|\Pi_{{\bf{h}}_1}{\bf{h}}_2\right\|}+\sqrt{1-\bar{x}^2}\frac{\Pi_{{\bf{h}}_1^{*}}^{\bot}{\bf{h}}_2^{*}}{\left\|\Pi_{{\bf{h}}_1}^{\bot}{\bf{h}}_2\right\|},
\end{align}
where $\bar{x}$ is a real number in $[0,1]$.
\end{proposition}
\proof Although the objective function is in a different form compared to \cite{E.A.Jorswieck}, the key idea to prove Proposition 1 is similar to the proof of Corollary 1 in \cite{E.A.Jorswieck}. Hence the proof is omitted here.\endproof

To this end, substituting (\ref{optw2}) into $(\ref{eqn:rate0})$, the original optimization problem $P0$ can be converted to
\begin{align}\label{P2}
P2:&\underset{\bar{x}, \tau}\max~~\frac{1-\tau}{2}\log_2\left(A_0(a\bar{x})^2+\frac{B_0(a\bar{x})^2\left(b\bar{x}+c\sqrt{1-\bar{x}^2}\right)^2}{C_0(a\bar{x})^2+D_0\left(b\bar{x}+c\sqrt{1-\bar{x}^2}\right)^2+1}\right)\\
& \begin{array}{r@{\quad}r@{}l@{\quad}l}
\mathrm{s.t.}& \bar{x} \in [0,1], \tau \in [0,1],
\end{array}
\end{align}
where $a=\frac{\left|{\bf{h}}_1^{T}\Pi_{{\bf{h}}_1^{*}}{\bf{h}}_2^{*}\right|}{\left\|\Pi_{{\bf{h}}_1}{\bf{h}}_2\right\|}$, $b=\left\|\Pi_{{\bf{h}}_1}{\bf{h}}_2\right\|$, $c=\left\|\Pi_{{\bf{h}}_1}^{\bot}{\bf{h}}_2\right\|$.

Since the objective function of the optimization problem $P2$ is a non-convex function w.r.t. $\bar{x}$ and $\tau$, a two-dimensional search method is required to obtain the optimal solution, which is nevertheless time-consuming. Motivated by this, in the following section, we propose a simple suboptimal design.

\section{Suboptimal Design}\label{section:suboptimal beamformer}
In this section, we propose suboptimal design method for the energy beamforming vector ${\bf w}$ and time split $\tau$. The major difficulty in the optimization problem $P0$ is the coupled relationship between ${\bf w}$ and $\tau$. To overcome this problem, we first devise a tight upper bound for the achievable throughput. By doing so, the optimization of ${\bf w}$ and $\tau$ can be decoupled, and closed-form expressions for the optimal energy beamforming vector ${\bf w}$ and time split $\tau$ can be obtained.

Noticing that $B_0=C_0*D_0$, it is easy to show that
\begin{align}\label{upper bound}
\frac{B_0\left|{\bf{h}}_1^T {\bf{w}}\right|^2\left|{\bf{h}}_2^T {\bf{w}}\right|^2}{C_0\left|{\bf{h}}_1^T {\bf{w}}\right|^2+D_0\left|{\bf{h}}_2^T {\bf{w}}\right|^2+1}\leq \min\left(C_0\left|{\bf{h}}_1^T {\bf{w}}\right|^2,D_0\left|{\bf{h}}_2^T {\bf{w}}\right|^2\right).
\end{align}

The above upper bound has been widely adopted in the literature in the analysis and design of AF relaying systems since it has been demonstrated that the upper bound remains sufficiently tight over the entire SNR range \cite{S.Ikki}. Hence, instead of maximizing the exact achievable throughput $R$, we focus on the maximization of the throughput upper bound.

Define $A=\frac{\left\|{\bf{h}}_1\right\|^2}{d_1^{2\alpha}}$, $C=\frac{\left|h_3\right|^2}{d_1^\alpha d_3^\alpha}$ and $D=\frac{\left\|{\bf{h}}_2\right\|^2}{d_2^{2\alpha}}$. Then, the original optimization problem $P0$ becomes
\begin{align}\label{P3_original}
P3:&\underset{{\bf{w}}, \tau}\max~~R^{u}=\frac{1-\tau}{2}\log_2\left(1+\frac{2\eta\rho\tau}{1-\tau}\left(A\left|{\bf{h}}_1^T {\bf{w}}\right|^2+\min\left(C\left|{\bf{h}}_1^T {\bf{w}}\right|^2,D\left|{\bf{h}}_2^T {\bf{w}}\right|^2\right)\right)\right)\\
&\mathrm{s.t.} \quad \left\|{\bf{w}}\right\|^2=1, \tau \in [0,1].
\end{align}

According to \cite{C.Zhong0}, the optimization problem $P3$ can be solved via sequential optimization of $\bf{w}$ and $\tau$. Hence, we first consider the optimization of ${\bf w}$.
\subsection{Energy Beamforming Vector Design}
It is easy to observe that, in terms of energy beamforming vector ${\bf w}$, the optimum $R^u$ is achieved when the term $\gamma\triangleq\left(A\left|{\bf{h}}_1^T {\bf{w}}\right|^2+\min\left(C\left|{\bf{h}}_1^T {\bf{w}}\right|^2,D\left|{\bf{h}}_2^T {\bf{w}}\right|^2\right)\right)$ is maximized. Following the same method as in the optimization problem $P1$, it can be shown that the optimal energy beamforming vector ${\bf w}$ admits the same form as in (\ref{optw2}). Therefore, the optimal ${\bf w}$ can be characterized by a single parameter $\bar{x}$, which is the solution to the following optimization problem:
\begin{align}\label{P3_0}
P4:&\underset{\bar{x}}\max\left\{Aa^2\bar{x}^2+\min\left\{Ca^2\bar{x}^2,D\left(b\bar{x}+c\sqrt{1-\bar{x}^2}\right)^2\right\}\right\}\\
&\begin{array}{r@{\quad}r@{}l@{\quad}l}
\mathrm{s.t.}&\bar{x} \in [0, 1],
\end{array}
\end{align}
which can be further transformed as
\begin{align}\label{P4}
P5:&\underset{\bar{x}}\max\left\{\min\left\{\left(Aa^2+Ca^2\right)\bar{x}^2,\left(Aa^2+D\left(b^2-c^2\right)\right)\bar{x}^2+2Dbc\bar{x}\sqrt{1-\bar{x}^2}+Dc^2\right\}\right\}\\
&\begin{array}{r@{\quad}r@{}l@{\quad}l}
\mathrm{s.t.}&\bar{x} \in [0, 1].
\end{array}
\end{align}
In order to solve $P5$, we find it convenient to give a separate treatment of three different scenarios according to the value of $\left(Aa^2+D\left(b^2-c^2\right)\right)\bar{x}^2$ as follows:

\subsection*{Scenario 1: $Aa^2+D\left(b^2-c^2\right)=0$}\label{subsection£ºthree case}
When $Aa^2+D\left(b^2-c^2\right)=0$ and the optimization problem $P5$ reduces to
\begin{align}
P6:&\underset{\bar{x}}\max\left\{\min\left\{\left(Aa^2+Ca^2\right)\bar{x}^2,2Dbc\bar{x}\sqrt{1-\bar{x}^2}+Dc^2\right\}\right\}\nonumber\\
&\begin{array}{r@{\quad}r@{}l@{\quad}l}
\mathrm{s.t.}&\bar{x} \in [0, 1].
\end{array}
\end{align}
Then, the optimal solution can be described in the following theorem:

\begin{theorem}\label{theorem k=0}
When $Aa^2+D\left(b^2-c^2\right)=0$, the solution of optimization problem $P5$ is given by
\begin{align}\label{item1and2}
\bar{x}=\left\{
\begin{array}{lcl}
1 &Aa^2+Ca^2 \leqslant Dc^2,\\
\frac{c\sqrt{D}}{\sqrt{\left(a\sqrt{C}-b\sqrt{D}\right)^2+Dc^2}} &Dc^2 < Aa^2+Ca^2 < 2\left(bc+c^2\right)D,\\
\frac{1}{\sqrt{2}} &Aa^2+Ca^2 \geqslant 2\left(bc+c^2\right)D.
\end{array}\right.
\end{align}
and the corresponding outcome of the objective function $\gamma$ is given by
\begin{align}\label{max gamma}
\gamma_{max}=\left\{
\begin{array}{lcl}
Aa^2+Ca^2 &\bar{x}=1,\\
\frac{D(A+C)a^2c^2}{\left(a\sqrt{C}-b\sqrt{D}\right)^2+Dc^2} &\bar{x}=\frac{d\sqrt{D}}{\sqrt{\left(a\sqrt{C}-b\sqrt{D}\right)^2+Dc^2}},\\
\left(bc+c^2\right)D &\bar{x}=\frac{1}{\sqrt{2}}.
\end{array}\right.
\end{align}
\end{theorem}
\proof Let us define $f_1(\bar{x})\triangleq \left(Aa^2+Ca^2\right)\bar{x}^2$ and $f_2(\bar{x}) \triangleq 2Dbc\bar{x}\sqrt{1-\bar{x}^2}+Dc^2$. Then $f_1(\bar{x})$ is a monotonically increasing function w.r.t. $\bar{x}$, and $f_2(\bar{x})$ is a concave function w.r.t $\bar{x}$, achieving the maximum value $\left(bc+c^2\right)+Dc^2$ at point $\bar{x}=\frac{1}{\sqrt{2}}$.
\begin{figure}
  \centering
  \subfigure[Case 1]{\label{fig:k=0,case1}\includegraphics[scale=0.5]{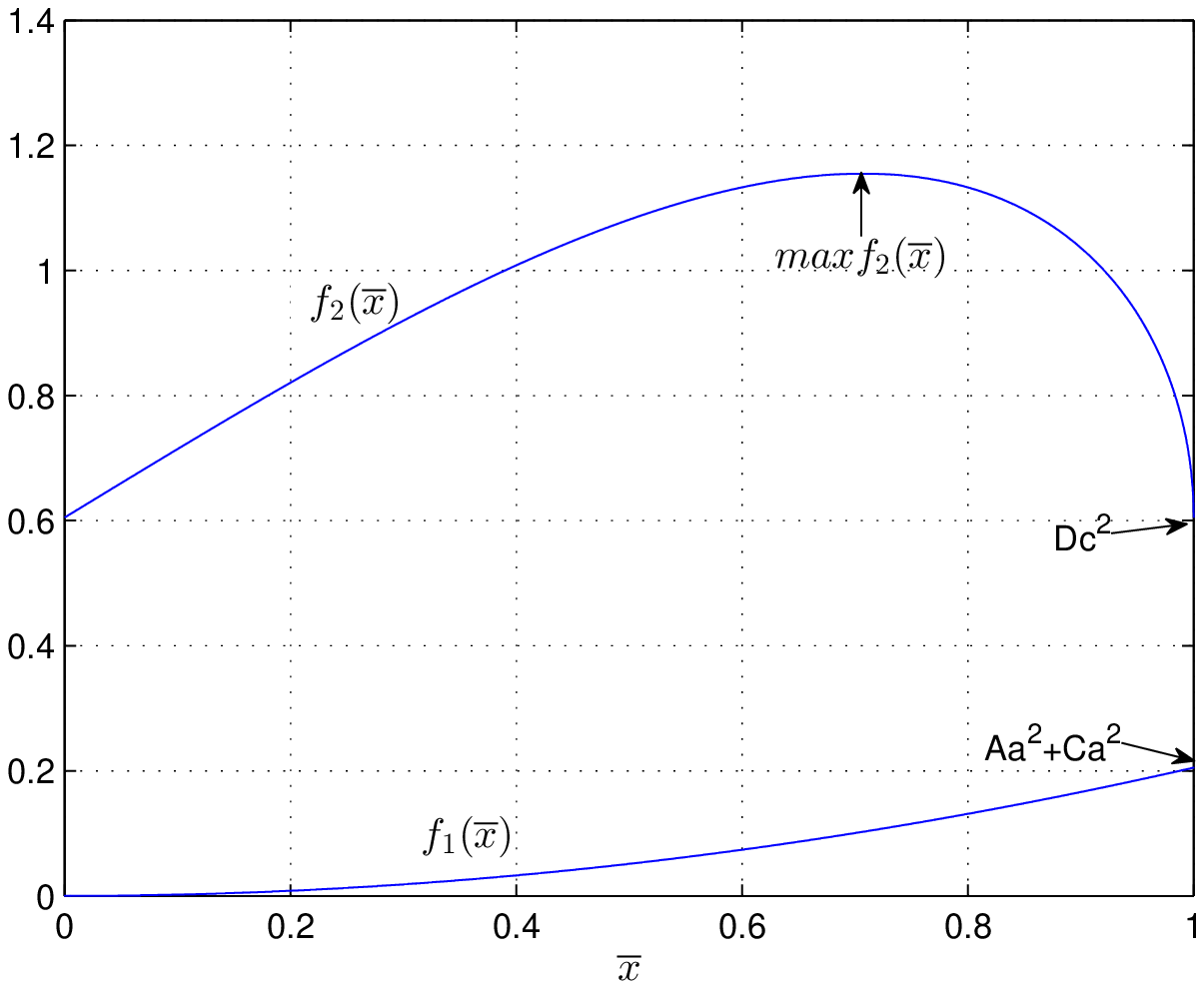}}
  \subfigure[Case 2]{\label{fig:k=0,case2}\includegraphics[scale=0.5]{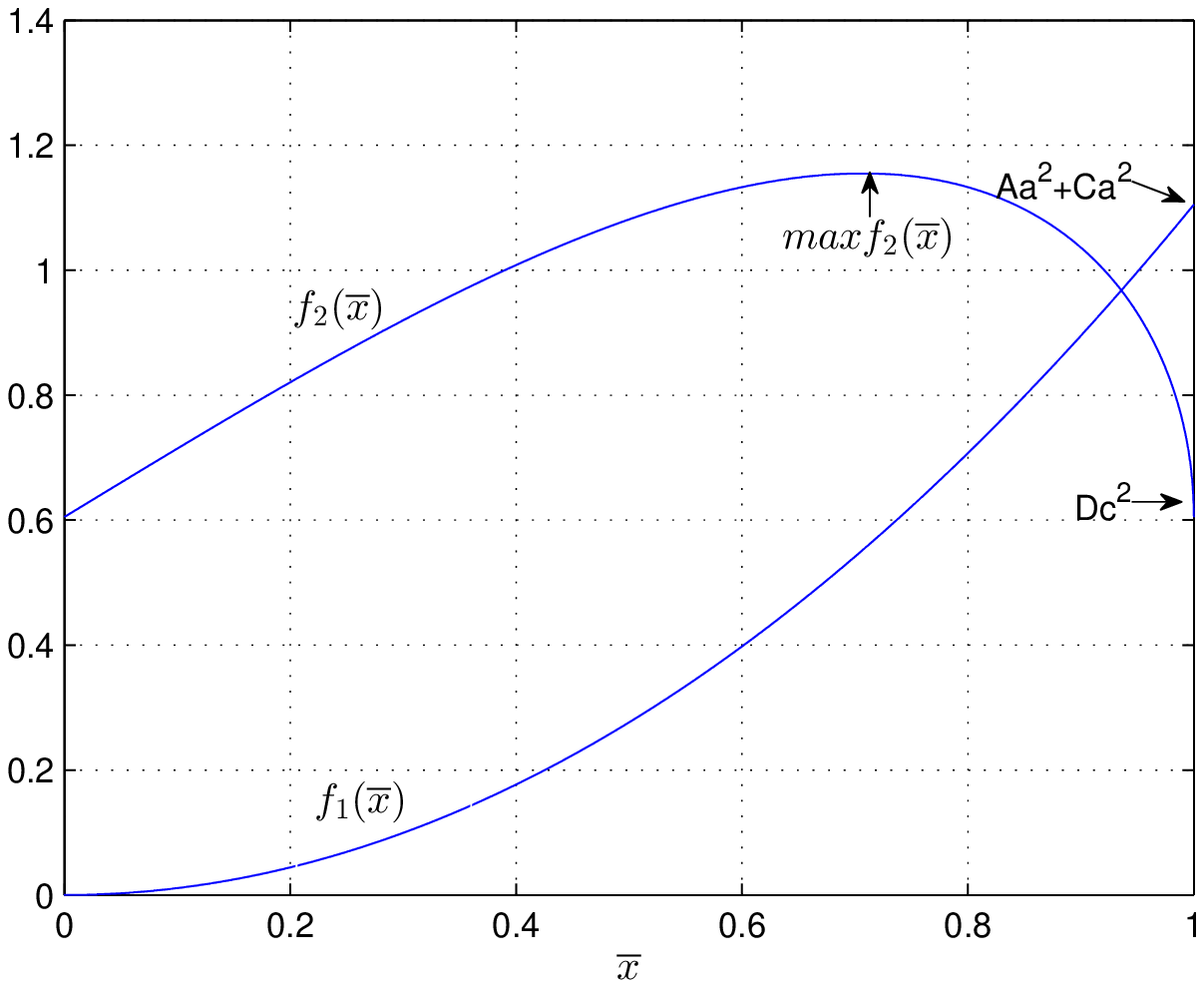}}
  \subfigure[Case 3]{\label{fig:k=0,case3}\includegraphics[scale=0.5]{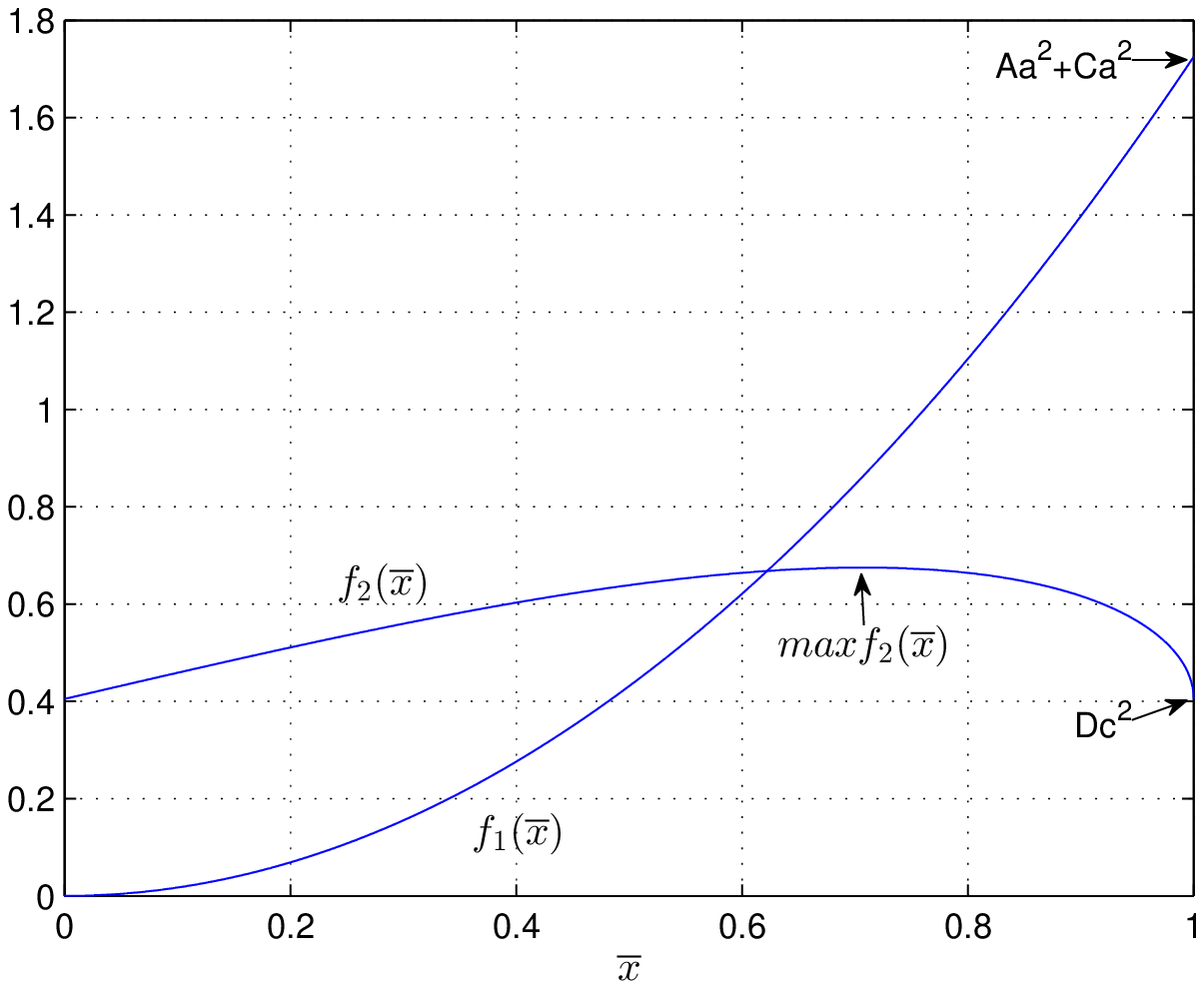}}
  \caption{Illustration of the three different cases.}\label{fig:k=0,three cases}
\end{figure}

As shown in Fig. \ref{fig:k=0,three cases}, there are three different cases to be considered:
\begin{itemize}
\item Case 1:
In this case, $Dc^2 \geqslant Aa^2+Ca^2$ and there is no intersecting point between the two curves, which implies that $f_2(\bar{x})$ is always greater than $f_1(\bar{x})$, i.e., $f_1(\bar{x}) \leqslant f_2(\bar{x})$ for all $\bar{x}$. Hence, the objective function $\gamma$ attains its maximum value at the point where $f_1(\bar{x})$ is maximized, i.e., $\bar{x}=1$, hence $\gamma_{max}=Aa^2+Ca^2$.

\item Case 2:
In this case, $Dc^2 < Aa^2+Ca^2 \leq 2\left(bc+c^2\right)D$, and there is a unique intersecting point between the two curves. A close observation reveals that the cross point appears after $f_2(\bar{x})$ reaching its maximum. Hence, the maximum value of the objective function $\gamma$ is attained at the cross point $f_1(\bar{x}) = f_2(\bar{x})$, i.e., $\bar{x}=\frac{d\sqrt{D}}{\sqrt{\left(a\sqrt{C}-b\sqrt{D}\right)^2+Dc^2}}$, and $\gamma_{max}=\frac{D(A+C)a^2c^2}{\left(a\sqrt{C}-b\sqrt{D}\right)^2+Dc^2}$.
\item Case 3:
In this case, $Aa^2+Ca^2 > 2\left(bc+c^2\right)D$, and the two curves also intersect at a unique point. However, different from Case 2, the cross point appears before $f_2(\bar{x})$ attaining its maximum value. Therefore, the maximum value of the objective function $\gamma$ is achieved at the point where $f_2(\bar{x})$ is maximized, i.e., $\bar{x}=\frac{1}{\sqrt{2}}$, hence $\gamma_{max}=\left(bc+c^2\right)D$.
\end{itemize}
\endproof

\subsection*{Scenario 2: $Aa^2+D\left(b^2-c^2\right)>0$}
We now turn to the case $Aa^2+D\left(b^2-c^2\right)>0$, and we have the following key result:
\begin{theorem}\label{theorem:case2}
When $Aa^2+D\left(b^2-c^2\right)> 0$, the solution of optimization problem $P6$ , i.e., $\bar{x}$, and the corresponding outcome of the objective function, i.e., $\gamma_{max}$, are given in (\ref{max xbar}) and (\ref{max gamma1}) respectively, shown at the top of the next page.

\end{theorem}

\proof
Let us define
\begin{align}
f_3(\bar{x}) \triangleq \left(Aa^2+D\left(b^2-c^2\right)\right)\bar{x}^2+2Dbc\bar{x}\sqrt{1-\bar{x}^2}+Dc^2.
\end{align}
Then, it is easy to show that the first derivative of $f_3(\bar{x})$, $f_3^{'}(\bar{x})=0$ has a unique root $\hat{x}=\sqrt{\frac{1}{2}+\frac{k^2}{2\sqrt{k^2+k^4}}}$ in $[0,1]$. Moreover, we have $f_3(\hat{x})>f_3(0)$ and $f_3^{''}(\hat{x})\neq0$ which implies $f_3(\bar{x})$ is an increasing function in the range $[0,\hat{x}]$ and a decreasing function in the range $[\hat{x},1]$.

Now, let $\Delta(\bar{x})$ denotes the difference of $f_3(\bar{x})$ and $f_1(\bar{x})$, we have
\begin{align}
\Delta(\bar{x})=\left(D\left(b^2-c^2\right)-Ca^2\right)\bar{x}^2+2Dbc\bar{x}\sqrt{1-\bar{x}^2}+Dc^2.
\end{align}
When $\left(D\left(b^2-c^2\right)-Ca^2\right)\leq 0$, it is easy to show that $\Delta(\bar{x})$ is a concave function. When $\left(D\left(b^2-c^2\right)-Ca^2\right)> 0$, then $\Delta(\bar{x})$ is similar to $f_3(\bar{x})$, i.e., it first increases, and then decreases. Since $\Delta(0)=Dc^2>0$ and $\Delta(1)=Db^2-Ca^2$, $\Delta(\bar{x})=0$ has at most one root in [0,1], which implies that $f_3(\bar{x})$ and $f_1(\bar{x})$ have at most one intersecting point in $[0,1]$. To this end, the problem can be solved by following the same approach as in Scenario 1.\endproof

\begin{figure*}[!t]
\begin{align}\label{max xbar}
\bar{x}=\left\{
\begin{array}{lcl}
1 &Aa^2+Ca^2 \leqslant 2Dbck+Dc^2,\\
\frac{D(A+C)a^2c^2}{\left(a\sqrt{C}-b\sqrt{D}\right)^2+Dc^2} &2DbcK+Dc^2<Aa^2+Ca^2<f_1(k),\\
\sqrt{\frac{1}{2}+\frac{k^2}{2\sqrt{k^2+k^4}}} &Aa^2+Ca^2 \geqslant f_1(k),
\end{array}\right.
\end{align}

where $k=\frac{Aa^2+D\left(b^2-c^2\right)}{2Dbc}$ and $f_1(k)=\frac{Dbc\left(k+\sqrt{\frac{1}{1+k^2}}+\frac{k^3}{\sqrt{k^2+k^4}}\right)+Dc^2}{\left(\frac{1}{2}+\frac{k^2}{2\sqrt{k^2+k^4}}\right)}$.
\begin{align}\label{max gamma1}
\gamma_{max}=\left\{
\begin{array}{lcl}
Aa^2+Ca^2 &\bar{x}=1,\\
\frac{D(A+C)a^2c^2}{\left(a\sqrt{C}-b\sqrt{D}\right)^2+Dc^2} &\bar{x}=\frac{d\sqrt{D}}{\sqrt{\left(a\sqrt{C}-b\sqrt{D}\right)^2+Dc^2}},\\
Dbc\left(k+\sqrt{\frac{1}{1+k^2}}+\frac{k^3}{\sqrt{k^2+k^4}}\right)+Dc^2 &\bar{x}=\sqrt{\frac{1}{2}+\frac{k^2}{2\sqrt{k^2+k^4}}}.
\end{array}\right.
\end{align}
\hrule
\end{figure*}

%



\subsection*{Scenario 3: $Aa^2+D\left(b^2-c^2\right)<0$}
We now consider the third case $Aa^2+D\left(b^2-c^2\right)<0$, and we have the following key result:
\begin{theorem}\label{theorem:case3}
When $Aa^2+D\left(b^2-c^2\right)< 0$, the solution of optimization problem $P6$, i.e., $\bar{x}$, and the corresponding outcome of the objective function, i.e., $\gamma_{max}$ is given in (\ref{max xbar2}) and (\ref{max gamma2}) respectively, shown at the top of the next page.
\end{theorem}
\proof When $Aa^2+D\left(b^2-c^2\right)<0$, $f_3(\bar{x})$ is a concave function. Hence, the desired results can be obtained by using the same approach as in Scenario 1.\endproof


We can observe from (\ref{optw2})  that the optimal energy beamformer is a linear combination of the MRT and zero-forcing (ZF) beamformers. In a game-theoretic framework \cite{E.A.Jorswieck2}, the parameter $x$ can be interpreted as the level of ``selfishness'' of the user. For instance, when $x=1$, the H-AP acts in a completely selfish way by using the MRT solution. When $x=0$, the H-AP becomes completely altruistic by applying the ZF beamformer. Whether the H-AP performs in selfish or altruistic way depends on the channel qualities of the direct link and the relay link. In the case where the channel quality of the direct link is much better, the H-AP tends to perform in a more selfish way and more energy is allocated to the user, while when the channel quality of the relay link is superior, the H-AP would act more altruistic since higher achievable throughput can be obtained by allocating more energy to the relay.

\subsection{Time Split Optimization}
Having obtained the optimal energy beamforming vector and the corresponding maximum SNR $\gamma_{max}$, the remaining task is to optimize the time split parameter. To do so, we observe that the joint optimization problem $P3$ reduces to
\begin{align}
P7:&\underset{\tau}\max~~R^{u}=\frac{1-\tau}{2}\log_2\left(1+\frac{2\eta\rho\tau}{1-\tau}\gamma_{max}\right)\\
&\mathrm{s.t.} \quad \tau \in [0,1].
\end{align}

To this end, capitalizing on the results presented in \cite{G.L.Moritz,C.Zhong02}, the above optimization problem can be solved and we have:
\begin{corollary}\label{time split}
The optimal time split $\tau$ can be expressed as
\begin{align}
&\tau_{opt}=\frac{e^{W\left(\frac{\kappa-1}{e}\right)+1}-1}{\kappa-1+e^{W\left(\frac{\kappa-1}{e}\right)+1}},
\end{align}
where $W(x)$ is the Lambert W function, and $\kappa=2\eta\rho\gamma_{max}$.
\end{corollary}

\subsection{Asymptotic Large $N$ Regime}
Recent advances in the communication theory have demonstrated the remarkable performance gains by employing a large number of antennas at the base station, commonly referred to as massive MIMO systems \cite{T.L.Marzetta,H.Xie2,H.Xie3}. In the context of wireless powered communications, increasing the number of antennas provides the additional benefit of elevated power transfer efficiency \cite{X.Chen1}. Therefore, it is of great interest to look into the asymptotic large antenna regime, as we pursue below. And we have the following key result.

\begin{figure*}[!t]
\begin{align}\label{max xbar2}
\bar{x}=\left\{
\begin{array}{lcl}
1 &Aa^2+Ca^2 \leqslant 2Dbck+Dc^2,\\
\frac{D(A+C)a^2c^2}{\left(a\sqrt{C}-b\sqrt{D}\right)^2+Dc^2} &2DbcK+Dc^2<Aa^2+Ca^2<f_2(k),\\
\sqrt{\frac{1}{2}-\frac{k^2}{2\sqrt{k^2+k^4}}} &Aa^2+Ca^2 \geqslant f_2(k),
\end{array}\right.
\end{align}
where $k$ is given in Theorem \ref{theorem:case2} and $f_2(k)=\frac{Dbc\left(k+\sqrt{\frac{1}{1+k^2}}-\frac{k^3}{\sqrt{k^2+k^4}}\right)+Dc^2}{\left(\frac{1}{2}-\frac{k^2}{2\sqrt{k^2+k^4}}\right)}$.
\begin{align}\label{max gamma2}
\gamma_{max}=\left\{
\begin{array}{lcl}
Aa^2+Ca^2 &\bar{x}=1,\\
\frac{D(A+C)a^2c^2}{\left(a\sqrt{C}-b\sqrt{D}\right)^2+Dc^2} &\bar{x}=\frac{d\sqrt{D}}{\sqrt{\left(a\sqrt{C}-b\sqrt{D}\right)^2+Dc^2}},\\
Dbc\left(k+\sqrt{\frac{1}{1+k^2}}-\frac{k^3}{\sqrt{k^2+k^4}}\right)+Dc^2 &\bar{x}=\sqrt{\frac{1}{2}-\frac{k^2}{2\sqrt{k^2+k^4}}}.
\end{array}\right.
\end{align}
\hrule
\end{figure*}

\begin{theorem}\label{large N}
In the asymptotic large antenna regime, i.e., $N\rightarrow\infty$, the optimal energy beamforming vector is given by
\begin{align}
{\bf{w}}=\bar{x}\frac{{\bf{h}}_1^{*}}{\left\|{\bf{h}}_1\right\|}+\sqrt{1-\bar{x}^2}\frac{{\bf{h}}_2^{*}}{\left\|{\bf{h}}_2\right\|},
\end{align}
where
\begin{align}
\bar{x}=\left\{
\begin{array}{lcl}
1~~~~&A\left\|{\bf{h}}_1\right\|^2-D\left\|{\bf{h}}_2\right\|^2\geqslant0,\\
\sqrt{\frac{D\left\|{\bf{h}}_2\right\|^2}{C\left\|{\bf{h}}_1\right\|^2+D\left\|{\bf{h}}_2\right\|^2}}~~~~~&A\left\|{\bf{h}}_1\right\|^2-D\left\|{\bf{h}}_2\right\|^2<0,
\end{array}
\right.
\end{align}
and the corresponding outcome of the objective function $\gamma$ is given by

\begin{align}
\gamma_{max}=\left\{
\begin{array}{lcl}
A\left\|{\bf{h}}_1\right\|^2 &\bar{x}=1,\\
\frac{\left(A\left\|{\bf{h}}_1\right\|^2+C\left\|{\bf{h}}_1\right\|^2\right)D\left\|{\bf{h}}_2\right\|^2}{C\left\|{\bf{h}}_1\right\|^2+D\left\|{\bf{h}}_2\right\|^2} &\bar{x}=\sqrt{\frac{D\left\|{\bf{h}}_2\right\|^2}{C\left\|{\bf{h}}_1\right\|^2+D\left\|{\bf{h}}_2\right\|^2}}.
\end{array}\right.
\end{align}
\end{theorem}
\proof
In the asymptotic large antenna regime, the vectors ${\bf{h}}_1^{*}$ and ${\bf{h}}_2^{*}$ become orthogonal. Hence, according to \cite{E.A.Jorswieck}, the optimal energy beamforming vector can be rewritten as:
\begin{align}\label{optw}
{\bf{w}}_{opt}=\bar{x}\frac{{\bf{h}}_1^{*}}{\left\|{\bf{h}}_1\right\|}+\bar{x}_1\frac{{\bf{h}}_2^{*}}{\left\|{\bf{h}}_2\right\|},
\end{align}
where $\bar{x}$ and $\bar{x}_1$ are complex scalar numbers and satisfy $\left|\bar{x}\right|^2+\left|\bar{x}_1\right|^2=1$.

\begin{figure}
  \centering
  \subfigure[Case 1]{\label{fig:k=0,case11}\includegraphics[scale=0.5]{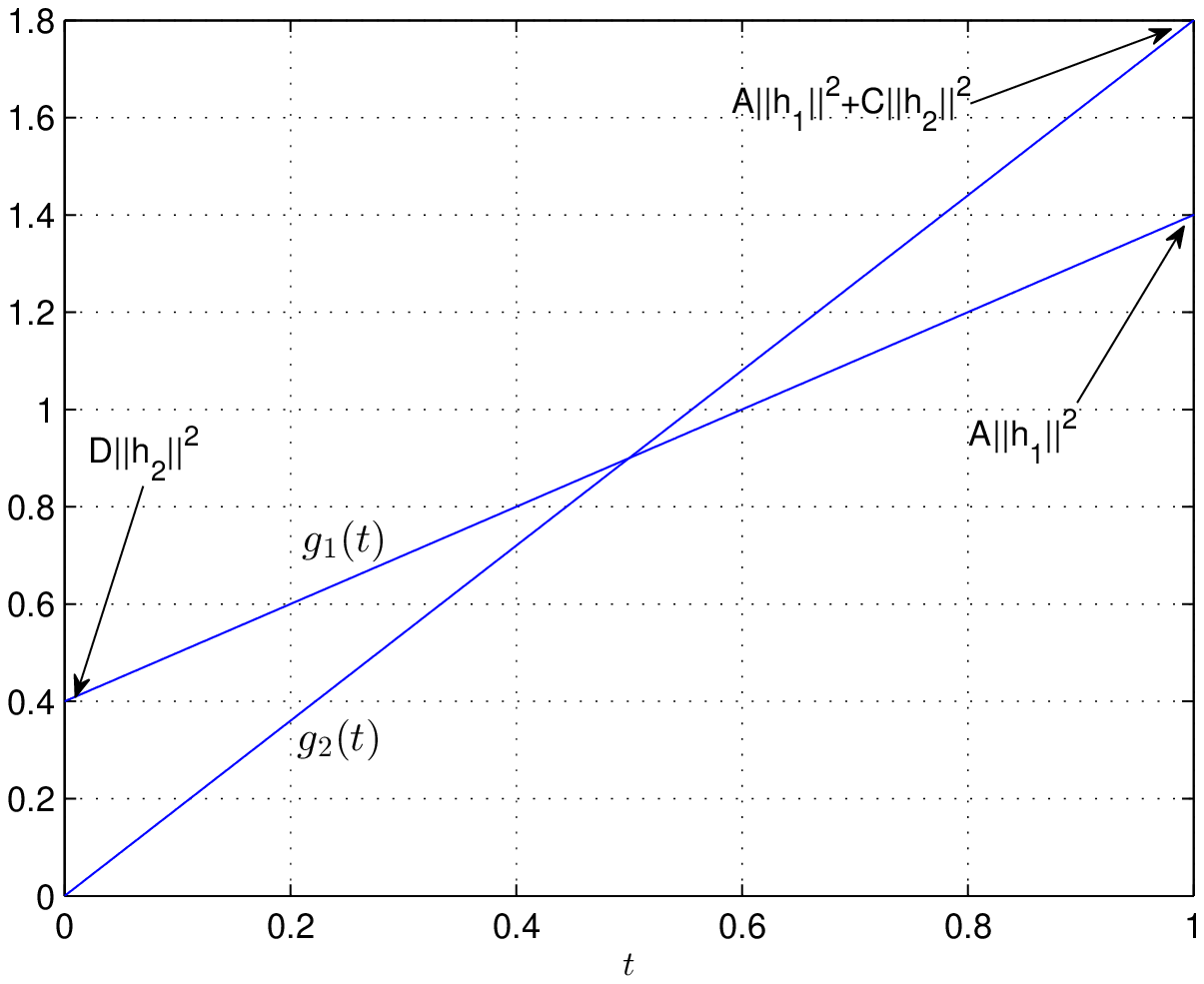}}
  \subfigure[Case 2]{\label{fig:k=0,case22}\includegraphics[scale=0.5]{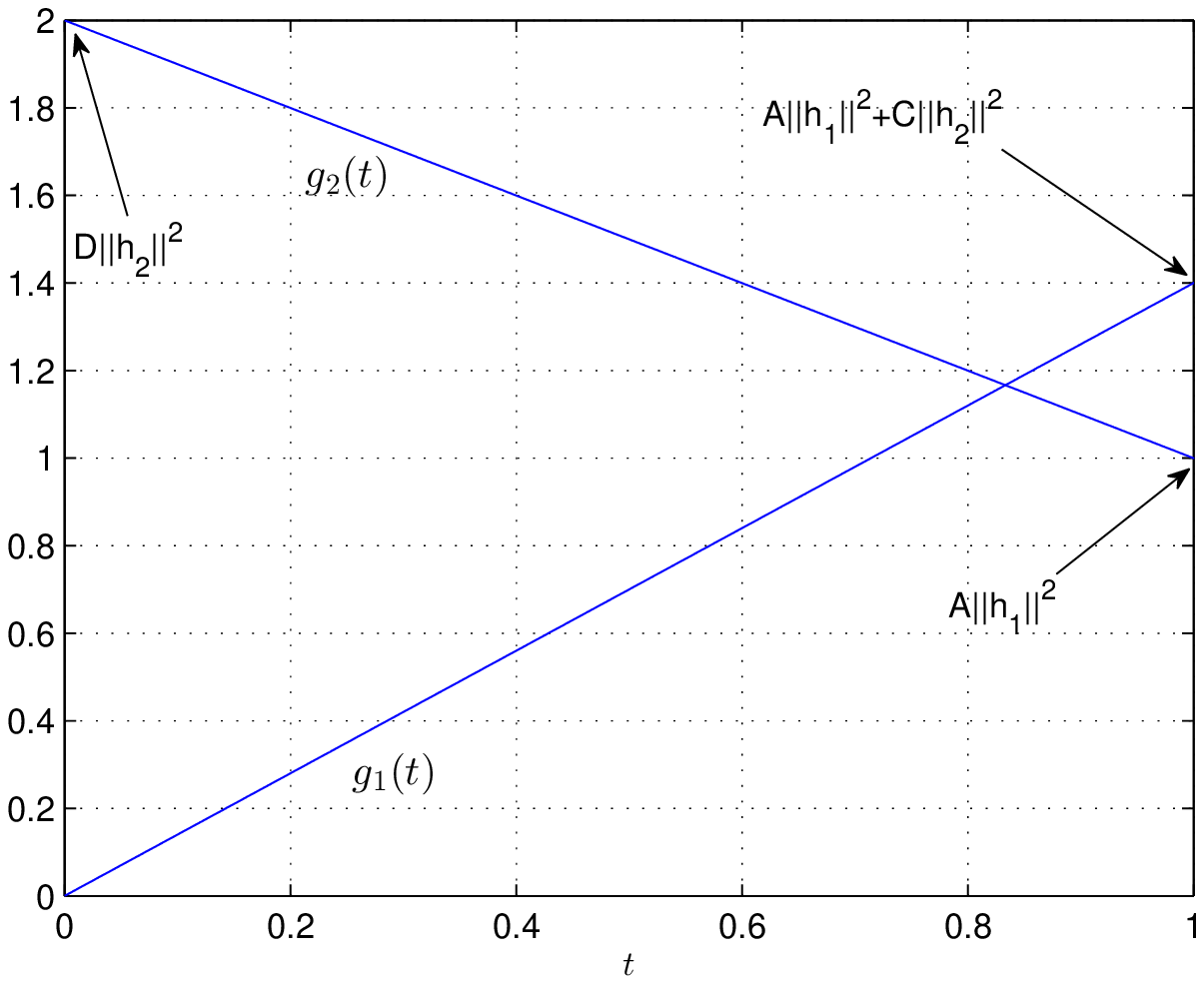}}
  \caption{Illustration of the two cases in asymptotic large $N$ regime.}\label{fig:largeNcases}
\end{figure}

Exploiting such orthogonality, the optimal beamforming vector can be obtained by solving the following optimization problem:
\begin{align}
P8:&\underset{\bar{x}}\max\left\{\min\left\{\left(A\left\|{\bf{h}}_1\right\|^2+C\left\|{\bf{h}}_1\right\|^2\right)\left|\bar{x}\right|^2,
\left(A\left\|{\bf{h}}_1\right\|^2-D\left\|{\bf{h}}_2\right\|^2\right)\left|\bar{x}\right|^2+D\left\|{\bf{h}}_2\right\|^2\right\}\right\}\\
&\mathrm{s.t.} ~~\bar{x}  \in \mathbb{C} \quad \text{and} \quad \left|\bar{x}\right| \in \left[0, 1\right].
\end{align}
First of all, noticing that the complex angle of $\bar{x}$ can be shifted by an arbitrary amount since it does not change the value of $\left|\bar{x}\right|$, without loss of generality, we take $\bar{x}$ to be a positive real number. Similarly, we can take $\bar{x}_1$ as a positive real number as well. Now define $t\triangleq \bar{x}^2$, $g_1(t)=\left(A\left\|{\bf{h}}_1\right\|^2+C\left\|{\bf{h}}_1\right\|^2\right)t$ and $g_2(t)=\left(A\left\|{\bf{h}}_1\right\|^2-D\left\|{\bf{h}}_2\right\|^2\right)t+D\left\|{\bf{h}}_2\right\|^2$. It is easy to observe that both $g_1(t)$ and $g_2(t)$ are linear functions w.r.t. $t$. In addition, it can be easily verified that $g_1(0)<g_2(0)$ and $g_1(1)>g_2(1)$. Therefore, there exists one unique intersecting point $\hat{t}$ in $[0, 1]$.  Then, two separate cases need to be considered as illustrated in Fig. \ref{fig:largeNcases}.

\begin{itemize}
\item Case 1: $A\left\|{\bf{h}}_1\right\|^2-D\left\|{\bf{h}}_2\right\|^2\geqslant0$, $\gamma_{max}$ is achieved at $t=1$, i.e., $\bar{x}=1$.
\item Case 2: $A\left\|{\bf{h}}_1\right\|^2-D\left\|{\bf{h}}_2\right\|^2<0$, $\gamma_{max}$ is achieved at the intersecting point of $g_1(t)$ and $g_2(t)$.
\end{itemize}
\endproof

We find that in the case $A\left\|{\bf{h}}_1\right\|^2-D\left\|{\bf{h}}_2\right\|^2\geqslant0$, i.e., $\frac{\left\|{\bf{h}}_1\right\|^2}{d_1^\alpha}\geqslant\frac{\left\|{\bf{h}}_2\right\|^2}{d_2^\alpha}$, the optimal beamforming strategy is to direct all the energy towards the user, i.e., the H-AP performs maximum ratio transmission along the direction of ${\bf{h}}_1$. This implies that, if the direct link is sufficiently strong, then contribution from the relay channel is negligible in the large $N$ regime.

\section{performance analysis}\label{section:performance analysis}
In this section, we present a detailed analysis on the achievable system performance. Due to the complicated expression of the energy beamforming vector derived in the previous section, further analysis is intractable. Therefore, in the following, we consider an intuitive energy beamforming scheme where the H-AP directs all the energy towards the user, i.e., ${\bf{w}}=\frac{{\bf{h}}_1^{*}}{\left\|{\bf{h}}_1\right\|}$. Specifically, we derive analytical expressions for two important performance metrics, namely, the outage probability and the average throughput. In addition, to gain further insights, we look into the high SNR regime, and derive a simple high SNR approximation for the outage probability.

When the H-AP applies the beamforming vector ${\bf{w}}=\frac{{\bf{h}}_1^{*}}{\left\|{\bf{h}}_1\right\|}$, the end-to-end SNR can be expressed as
\begin{align}\label{total SNR multi}
\gamma_{\sf m}=a_1\left\|{\bf{h}}_1\right\|^4
+\frac{{b_1c_1\left\|{\bf{h}}_1\right\|^2\left|h_3\right|^2\frac{\left|{\bf{h}}_2^T {\bf{h}}_1^{*}\right|^2}{\left\|{\bf{h}}_1\right\|^2}\left\|{\bf{h}}_2\right\|^2}}{{b_1\left\|{\bf{h}}_1\right\|^2\left|h_3\right|^2}+{c_1\frac{\left|{\bf{h}}_2^T {\bf{h}}_1^{*}\right|^2}{\left\|{\bf{h}}_1\right\|^2}\left\|{\bf{h}}_2\right\|^2}+1},
\end{align}
where $a_1=\frac{2\eta\tau\rho}{(1-\tau)d_1^{2\alpha}}$, $b_1=\frac{2\eta\tau\rho}{(1-\tau)d_1^\alpha d_3^{\alpha}}$ and $c_1=\frac{2\eta\tau\rho}{(1-\tau)d_2^{2\alpha}}$.

\subsection{Outage Probability}
Mathematically, the outage probability is defined as the probability of the instantaneous SNR falls below a pre-defined threshold $\gamma_{\sf th}$, i.e.,
\begin{align}\label{outage definition}
P_{\sf out}=P\left\{\gamma_{\sf m} < \gamma_{\sf th}\right\}.
\end{align}

\begin{theorem}\label{theorem outmulti}
With ${\bf{w}}=\frac{{\bf{h}}_1^{*}}{\left\|{\bf{h}}_1\right\|}$, the outage probability of the system can be expressed as
\begin{multline}\label{outage multi0}
P_{\sf out}=\int\limits_0^{\sqrt{\frac{\gamma_{\sf th}}{a_1}}}\left(1-e^{\frac{a_1}{b_1}y-\frac{\gamma_{\sf th}}{b_1y}}\right)\frac{y^{N-1}}{\Gamma\left(N\right)}
e^{-y}dy+\int\limits_0^{\sqrt{\frac{\gamma_{\sf th}}{a_1}}}\int\limits_{\frac{\gamma_{\sf th}-a_1y^2}{b_1y}}^{\infty}\left(1-2\sum_{m=0}^{N-1}\sum_{i=0}^{N-2}\binom{N-2}{i}\right.\\
\left.\times\frac{(-1)^i\left(N-1\right)}{m!} \chi^{i+1}\Gamma\left(m-2i-2,\sqrt{\chi}\right)\right)
e^{-\mu}d\mu\frac{y^{N-1}e^{-y}}{\Gamma\left(N\right)}dy,
\end{multline}
where $\chi=\frac{(b_1y\mu+1)(\gamma_{\sf th}-a_1y^2)}{b_1c_1y\mu+a_1c_1y^2-\gamma_{\sf th}c_1}$.

\proof See Appendix \ref{appendix:theorem outmulti}.
\endproof

\end{theorem}

While Theorem \ref{theorem outmulti} provides an efficient method for evaluating the exact outage probability of the system, this expression is quite complicated, and do not allow for easy extraction of useful insights. Motivated by this, we now look into the high SNR regime, and derive a simple approximation for the outage probability, which enables the characterization of the achievable diversity order.
\subsection{High SNR Approximation}
\begin{theorem}\label{theorem:highsnr}
With ${\bf{w}}=\frac{{\bf{h}}_1^{*}}{\left\|{\bf{h}}_1\right\|}$, in the high SNR regime, the outage probability can be approximated by
\begin{align}\label{eq:highsnr}
P_{\sf out}^{\infty}=\frac{2d_3^\alpha}{d_1^\alpha\Gamma\left(N\right)(N+1)(N-1)}\left[\frac{(1-\tau)d_1^{2\alpha}\gamma_{\sf th}}{2\eta\tau\rho}\right]^{\frac{N+1}{2}}.
\end{align}

\proof See Appendix \ref{appendix:theorem:highsnr}.
\endproof

\end{theorem}
We observe that the system achieves a diversity order of $\frac{N+1}{2}$ which is half of that in conventional cooperative relay systems with constant power supply for both the relay and user. The reason is that the received signal at the BS experiences twice channel fading. In addition, we find the impact of $d_2$ vanishes in the high SNR regime. The reason is that, the performance of the dual-hop relay link is limited by the weakest hop, and in the high SNR regime, the H-AP to relay link outperforms the relay to user link because of the available multiple antennas at the H-AP.

\subsection{Average Throughput}

In this subsection, we study the average throughput for the proposed system. Mathematically, the average throughput is defined as the expected value of the instantaneous mutual information, and it is given by
\begin{align}\label{throughput lowerbound}
C=\frac{1-\tau}{2}\mathbb{E}\left[\log_2\left(1+\gamma_{\sf m}\right)\right]=\frac{1-\tau}{2}\mathbb{E}\left[\log_2\left(1+\gamma_{us}+\frac{\gamma_{ur}\gamma_{rs}}{\gamma_{ur}+\gamma_{rs}+1}\right)\right],
\end{align}
where $\gamma_{us}=a_1\left\|{\bf{h}}_1\right\|^4$, $\gamma_{ur}=b_1\left\|{\bf{h}}_1\right\|^2\left|h_3\right|^2$ and $\gamma_{rs}=c_1\frac{\left|{\bf{h}}_2^T {\bf{h}}_1^{*}\right|^2}{\left\|{\bf{h}}_1\right\|^2}$.

Unfortunately, an exact evaluation of the average throughput is intractable due to the difficulty in obtaining the closed-form expression for the CDF of (\ref{total SNR}). In order to circumvent the issue, we now look into a tight approximation for the average throughput.
Noticing that $f\left(x, y\right)=\log_2\left(1+e^x+e^y\right)$ is a convex function w.r.t. $x$ and $y$, invoking the following result
\begin{align}\label{convex approximation}
\mathbb{E}\left[\log_2\left(1+\gamma_1+\gamma_2\right)\right]\geqslant \log_2\left(1+e^{\mathbb{E}\left(\ln{\gamma_1}\right)}+e^{\mathbb{E}\left(\ln{\gamma_2}\right)}\right),
\end{align}
we establish the following theorem:

\begin{theorem}\label{theorem:throughput}
With ${\bf{w}}=\frac{{\bf{h}}_1^{*}}{\left\|{\bf{h}}_1\right\|}$, the average throughput of the system can be lower bounded by
\begin{align}\label{lower bound multi}
C_{\sf low}=\frac{1-\tau}{2}\log_2\left(1+e^{m_1}+e^{\left[m_2+m_3-\ln\left(1+m_4+m_5\right)\right]}\right),
\end{align}
where $m_1=\ln a_1+2\psi(1)+2\sum_{m=1}^{N-1}\frac{1}{m}$, $m_2=\ln{b_1}+2\psi(1)+\sum_{m=1}^{N-1}\frac{1}{m}$, $m_3=\ln{c_1}+2\psi(1)+2\sum_{m=1}^{N-1}\frac{1}{m}-(N-1)\sum_{i=0}^{N-2}\binom{N-2}{i}(-1)^i\frac{1}{(i+1)^2}$, $m_4=\frac{b_1N(N-1)}{2}$ and $m_5=4c_1(N-1){\sum_{m=0}^{N-1}}\frac{1}{m!}{\sum_{i=0}^{N-2}}\\ \times {\binom{N-2}{i}}{(-1)^i}\frac{\Gamma(m+2)}{2i+4}$.

\proof See Appendix \ref{appendix:theorem:throughput}.
\endproof
Theorem \ref{theorem:throughput} presents a new lower bound for the average throughput of the system, which is quite tight across the entire SNR range as shown in the Section \ref{numerical result}, hence, providing an efficient means to evaluate the throughput without resorting to time-consuming Monte Carlo simulations.

\end{theorem}

\section{Numerical Results}\label{numerical result}
In this section, numerical results are presented to validate the analytical expressions as well as to illustrate the impact of key parameters on the system performance. Unless otherwise specified, the following set of parameters are used in the simulations: The carrier frequency is $5$ GHz, and the bandwidth is $20$ MHz, the noise power density is $-174$ dBm/Hz, the energy conversion efficiency is $\eta=0.5$, the path loss exponent is $\alpha=2.5$, and the number of antennas at the H-AP is $N=10$. Apart from the outage curves which are generated by averaging over $10^7$ independent channel realizations, all the other simulation results are obtained by averaging over $10^4$ independent channel realizations.

\begin{figure}[!ht]
  \centering
  {\includegraphics[scale=0.65]{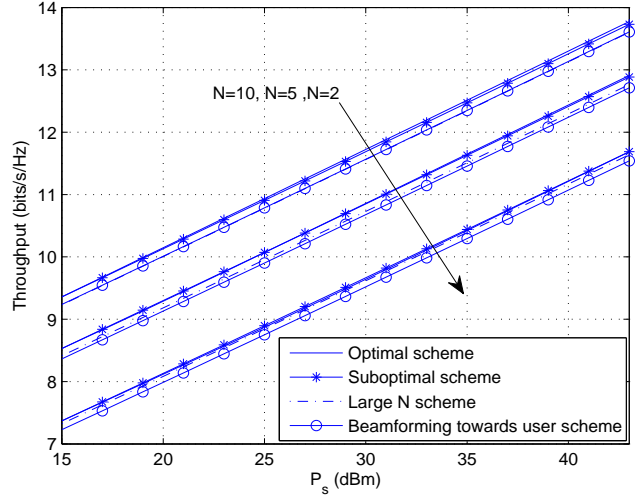}}
  \caption{Throughput comparison between the optimal and suboptimal schemes.}
  \label{fig:optbeam}
\end{figure}

%

Fig. \ref{fig:optbeam} illustrates the achievable throughput of the system with optimal and suboptimal schemes when $d_1=d_2=20$ m and $d_3 =2$ m. In particular, the curves associated with the optimal scheme are obtained via a two-dimensional search according to (\ref{P2}), the curves associated with the suboptimal scheme are plotted according to Theorem \ref{theorem k=0}, \ref{theorem:case2}, \ref{theorem:case3} and Corollary \ref{time split}, and the curves associated with the large $N$ approximation scheme are plotted according to Theorem {\ref{large N}} and the curves associated with beamforming towards user scheme are plotted according to ${\bf{w}}=\frac{{\bf{h}}_1^{*}}{\left\|{\bf{h}}_1\right\|}$. It can be readily observed that the performance gap between the optimal and suboptimal schemes is almost negligible across the entire SNR range of interest for different $N$, which indicates that the suboptimal scheme achieves almost the same performance as the optimal scheme. Therefore, the suboptimal design is preferable from a practical implementation point. In addition, we find that the suboptimal schemes always outperforms the beamforming towards user scheme. This is also expected because the design of suboptimal scheme exploits more CSI. Interestingly, the large $N$ approximation is accurate even for small $N$, i.e., $N=2$, and almost overlaps with the optimal scheme with moderate $N$, i.e., $N=10$. Moreover, we observe the intuitive beamforming towards user scheme achieves similar performance as the large $N$ approximation scheme. In addition, the achievable throughput increases monotonically as the number of antennas $N$ becomes larger, as expected. Please note, the suboptimal scheme and beamforming towards user scheme are both low-complexity designs. The suboptimal scheme attains near optimal performance, hence is attractive for practical implementation. While the beamforming towards user scheme serves as a good alternative in certain special case where the CSI of the relay link is unavailable at the H-AP.

\begin{figure}[!ht]
  \centering
  \subfigure[Impact of relay position when $P_s = 40$ dBm]{\label{fig:Impact of distance2}\includegraphics[scale=0.5]{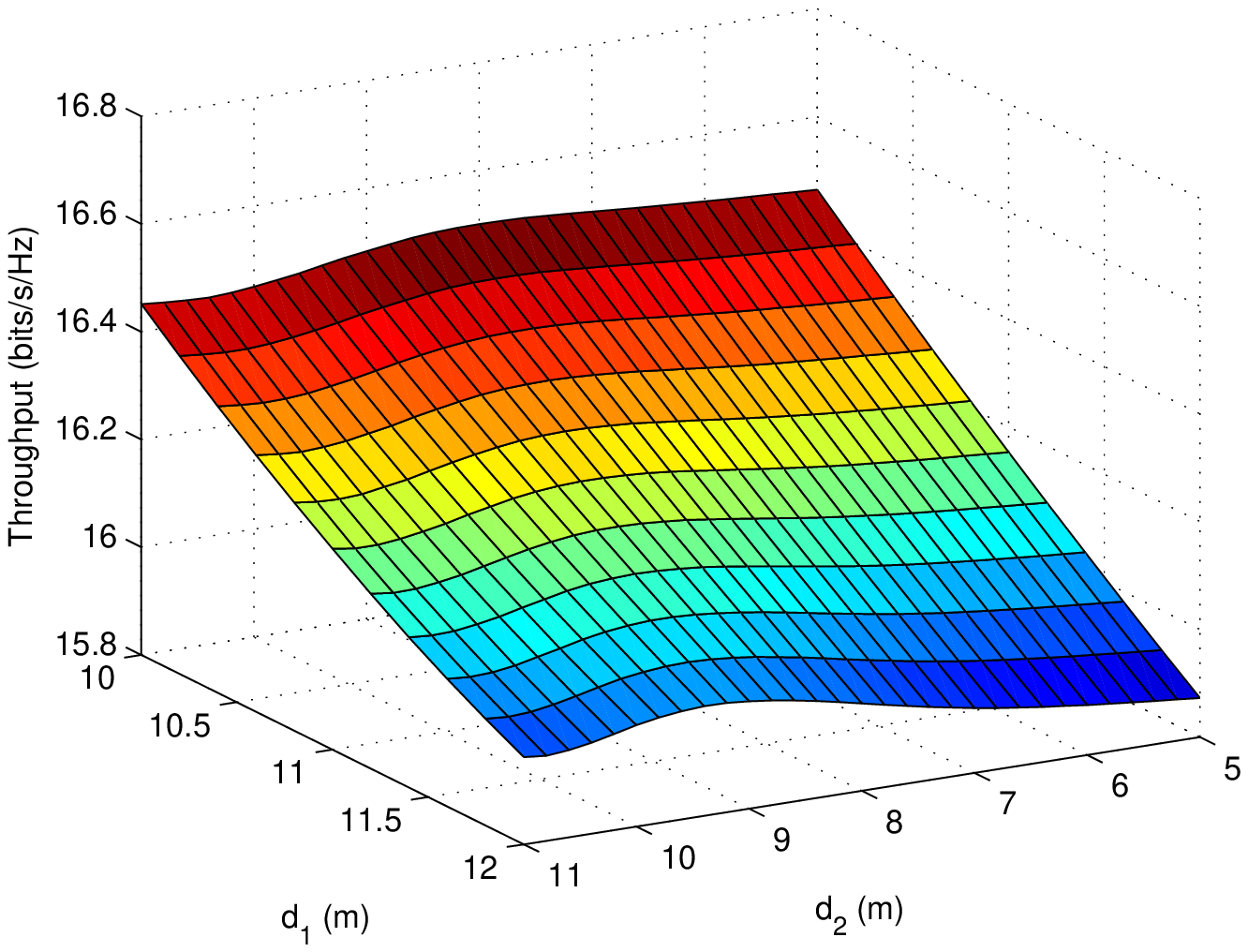}}
    \subfigure[Impact of $\rho$ on the relay position]{\label{fig:Impact of distance}\includegraphics[scale=0.5]{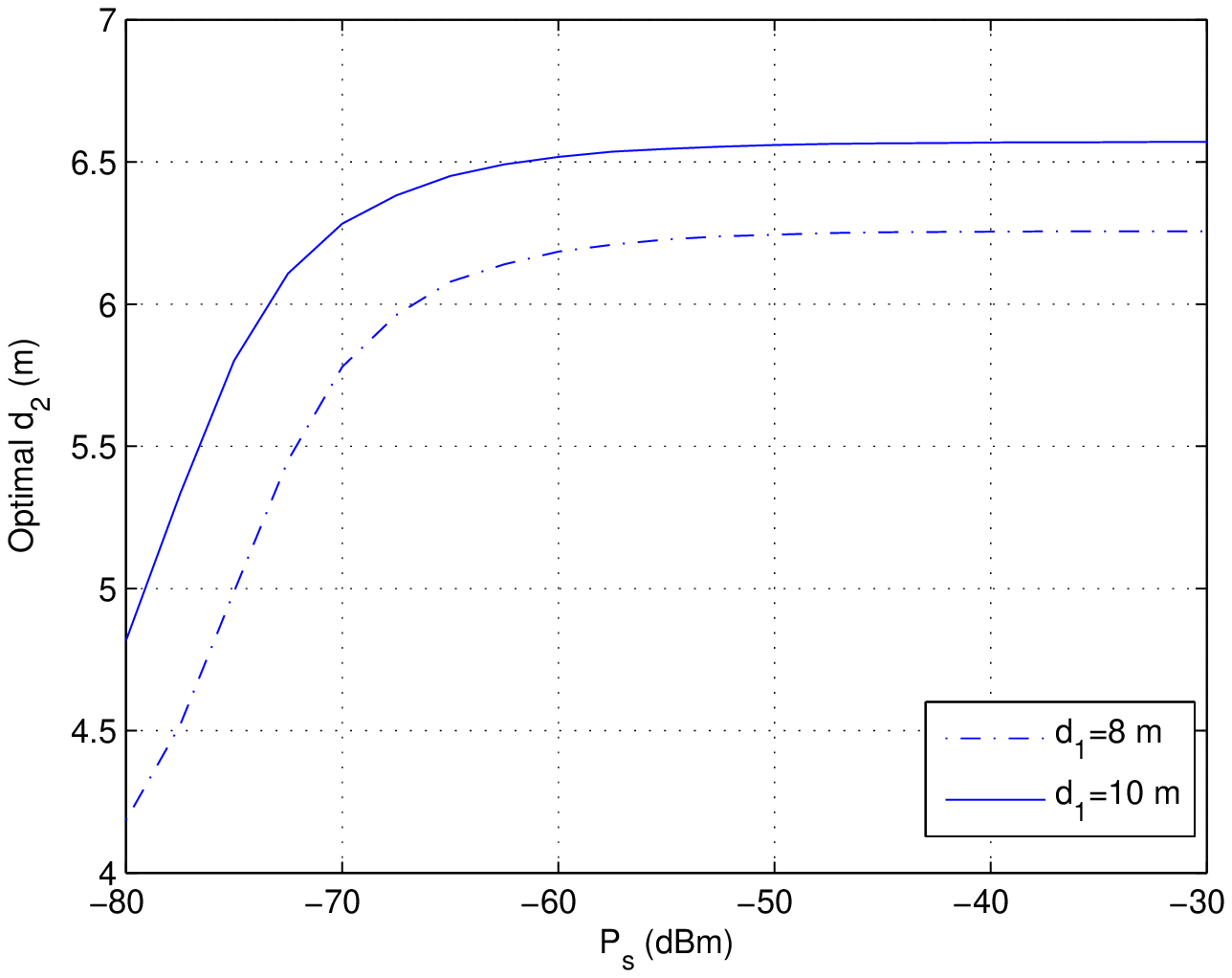}}
  \caption{Impact of node positions on the achievable throughput.}\label{fig:distance}
\end{figure}

Fig. \ref{fig:distance} examines the impact of the node locations on the achievable throughput with $d_2+d_3=12$ m and $N=20$. As shown in Fig. \ref{fig:Impact of distance2}, the impact of distance between the user and H-AP $d_1$ is significant, and the achievable throughput decreases monotonically when $d_1$ increases. This is intuitive since the increasing $d_1$ results in two-fold negative effects, i.e., reducing the amount of harvested energy at the user and degrading the quality of information transmission. Also, for each $d_1$, there exists an optimal relay position which maximizes the achievable throughput. It can be observed that, when $d_1$ increases, the optimal $d_2$ also increases. This can be explained as follows, when $d_1$ increases, less energy can be harvested by the user, hence, the relay has to move closer to the user to achieve a fine balance between the two hops. This behavior can also be observed in Fig. \ref{fig:Impact of distance}, where the optimal $d_2$ for $d_1=10$ m is larger than that for $d_1=8$ m. In addition, for fixed $d_1$, as $\rho$ increases, the optimal $d_2$ also increases, and gradually settles at sufficiently high SNRs. The reason is that, with multiple antennas at the BS, the benefit of increasing $\rho$ is more significant for the relay-BS link. As such, the optimal $d_2$ should increase in order to maintain the balance between the two hop channels, since the performance of the dual-hop relay link is limited by the weakest hop.

\begin{figure}[!ht]
  \centering
  {\includegraphics[scale=0.65]{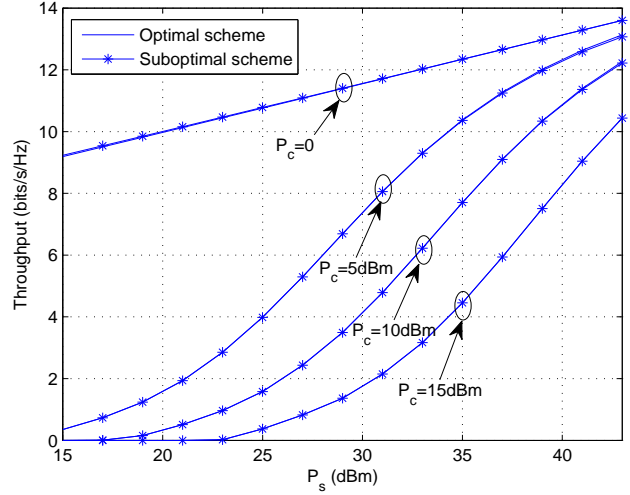}}
  \caption{Impact of circuit power consumption on the system performance when $d_1=20$ m and $d_2=d_3=15$ m.}
  \label{fig:circuit}
\end{figure}

{This work has assumed that the power required for circuit operation is not supplied by the harvested energy. However, it is still of interest to investigate the impact of circuit power consumption on the system performance when the circuitry is powered via energy harvesting, as illustrated in Fig. \ref{fig:circuit}, where a constant circuit power consumption $P_c$ is considered as in \cite{D.W.K.Ng}.\footnote{{Taking into account of circuit power consumption, the transmit power of the user and the relay during the information transmission phase become $P_{u}=\max\left\{0,\frac{2\eta\tau P_s\left|{\bf{h}}_1^T {\bf{w}}\right|^2}{(1-\tau)d_1^\alpha}-P_c\right\}$ and $P_{r}=\max\left\{0,\frac{2\eta\tau P_s\left|{\bf{h}}_2^T {\bf{w}}\right|^2}{(1-\tau)d_2^\alpha}-P_c\right\}$. Then, following the same method as in the optimization problem $P1$, it can be shown that the optimal energy beamforming vector ${\bf{w}}$ admits the same form as in (\ref{optw2}) in this scenario. Therefore, the curves associated with the optimal scheme and suboptimal scheme can be obtained via a two-dimensional search.}}
As can be readily observed, the performance gap is substantial in the low SNR regime, and the gap becomes larger as the circuit power consumption $P_c$ increases. This is because the amount of the harvested energy is too small to activate the circuitry in the low SNR regime, hence no information can be transmitted. In contrast, in the high SNR regime, the performance gap is significantly reduced. This observation suggests that it is necessary to have external power supply for the circuitry.}


\begin{figure}[!ht]
  \centering
  \subfigure[Impact of $N$ on the optimal $\tau$.]{\label{fig:optimal_tau_vs_ps}\includegraphics[scale=0.5]{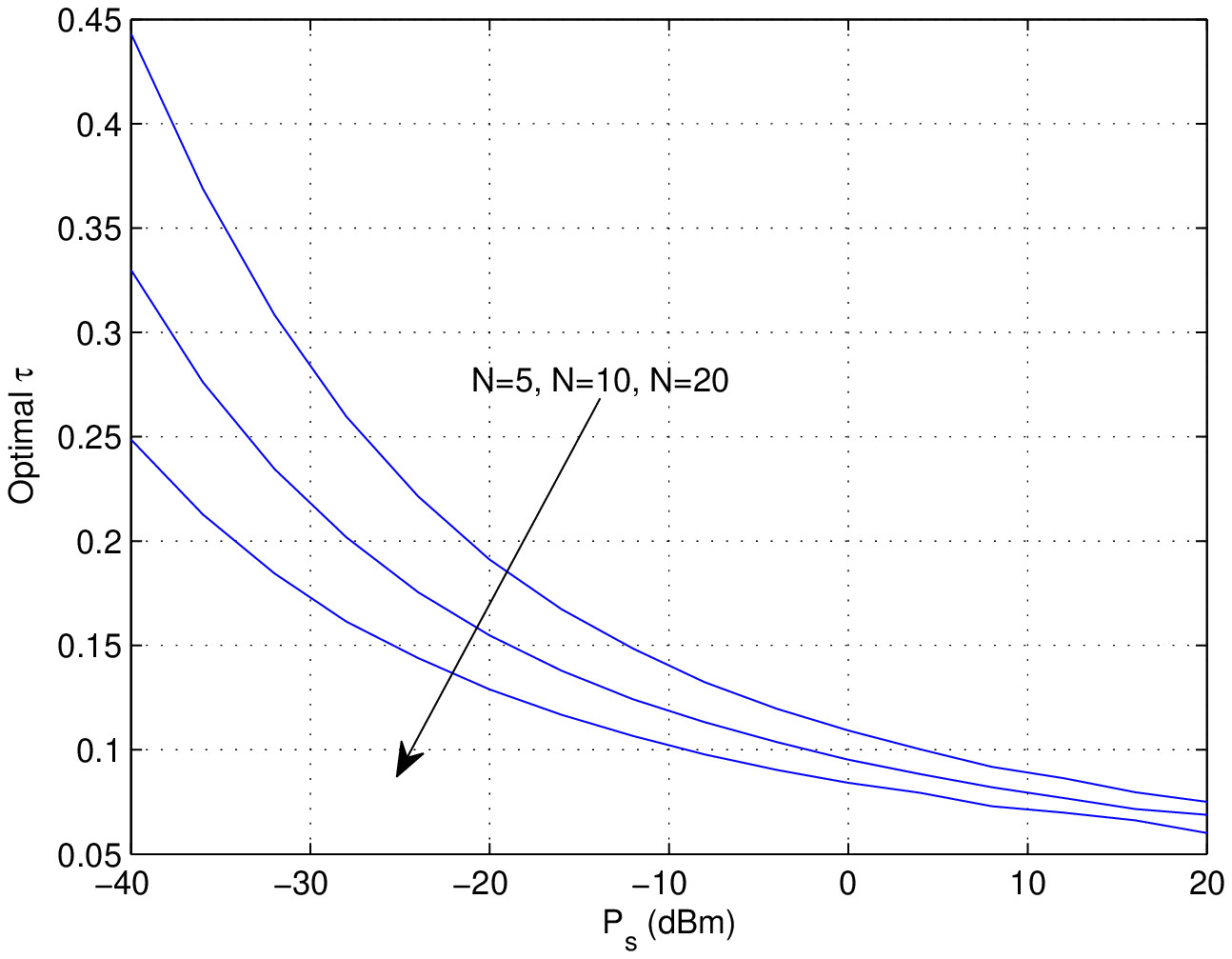}}
    \subfigure[Impact of direct link on the optimal $\tau$.]{\label{fig:optimal_tau_vs_ps_N}\includegraphics[scale=0.5]{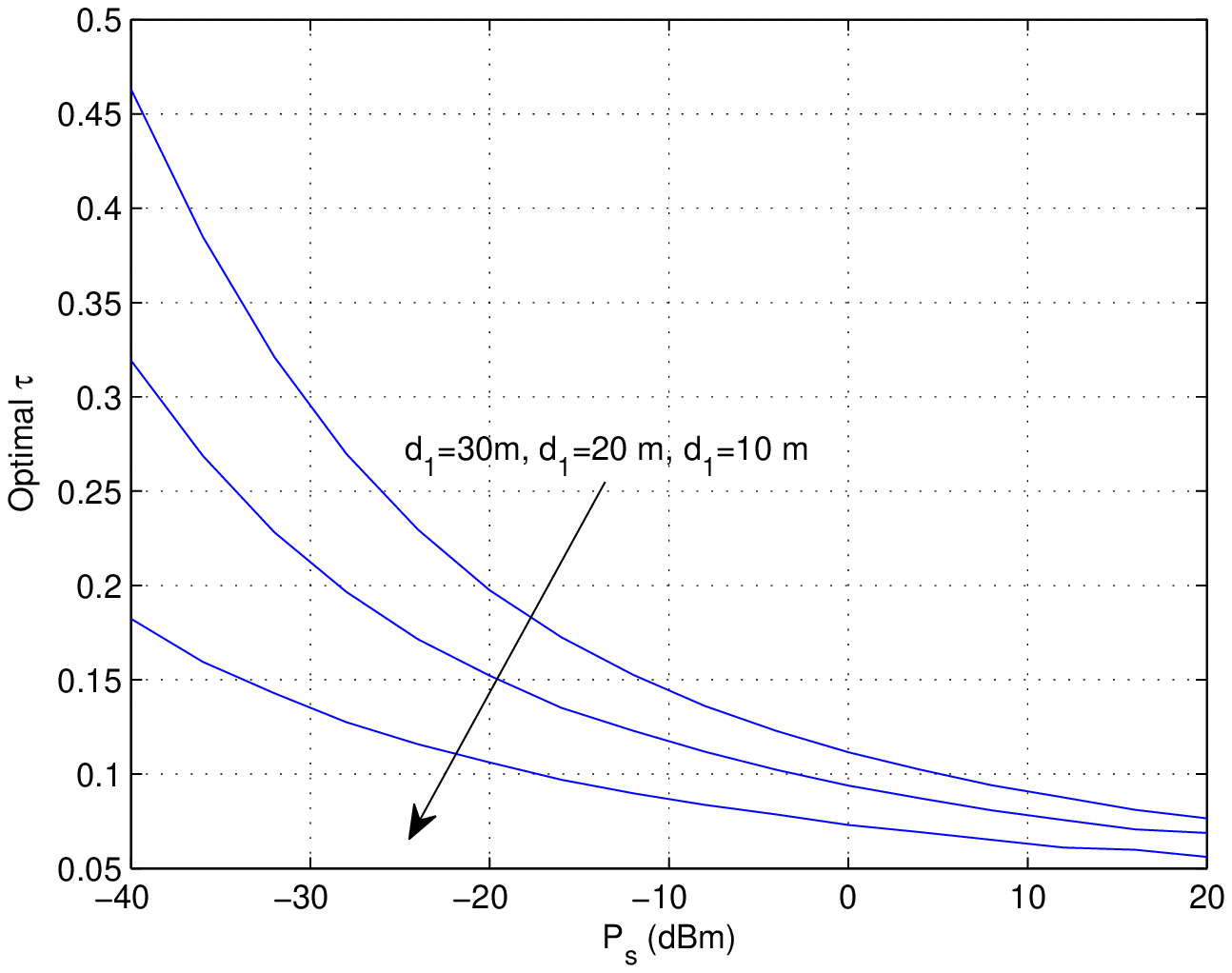}}
  \caption{Impact of antenna number and direct link on the optimal time split.}\label{fig:opttau}
\end{figure}

Fig. \ref{fig:opttau} illustrates the impact of $N$ and $d_1$ on the optimal time split $\tau$. As shown in Fig. \ref{fig:optimal_tau_vs_ps}, for all cases, the optimal $\tau$ decreases as $P_s$ increases. Also, increasing the number of antennas leads to a decrease of the optimal $\tau$. This is as expected since a large $N$ can significantly boost up the energy transfer efficiency, thereby reducing the energy harvesting time. Similarly, Fig. \ref{fig:optimal_tau_vs_ps_N} demonstrates the intuitive results that less energy harvesting time is required when the distance between the user and BS becomes smaller. In addition, it is observed that the impact of $N$ and $d_1$ on the optimal $\tau$ is more pronounced in the low $P_s$ regime.

%

\begin{figure}[!ht]
  \centering
  \subfigure[Impact of relay link on throughput performance.]{\label{fig:01}\includegraphics[scale=0.5]{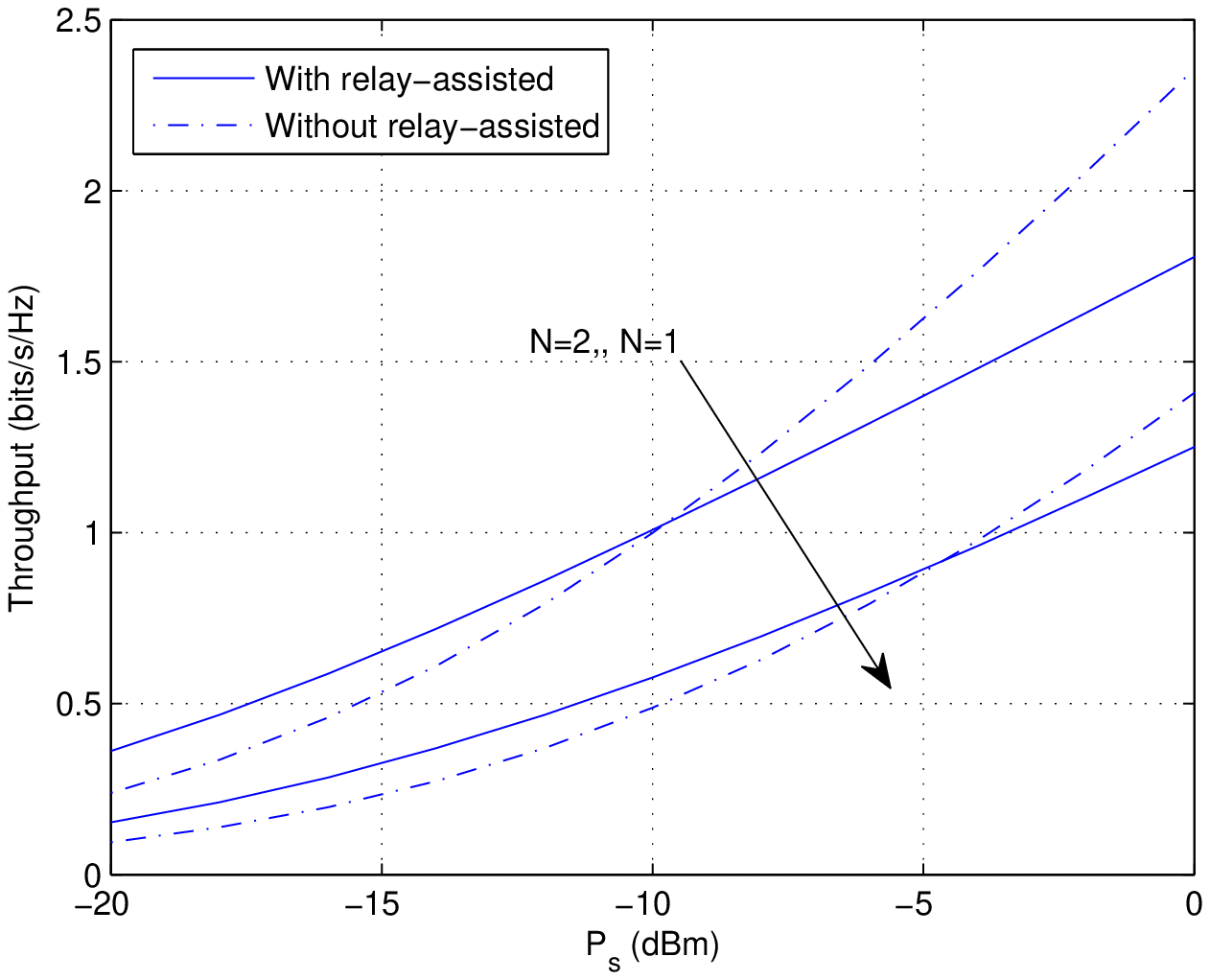}}
  \subfigure[Impact of relay link on outage performance.]{\label{fig:02}\includegraphics[scale=0.5]{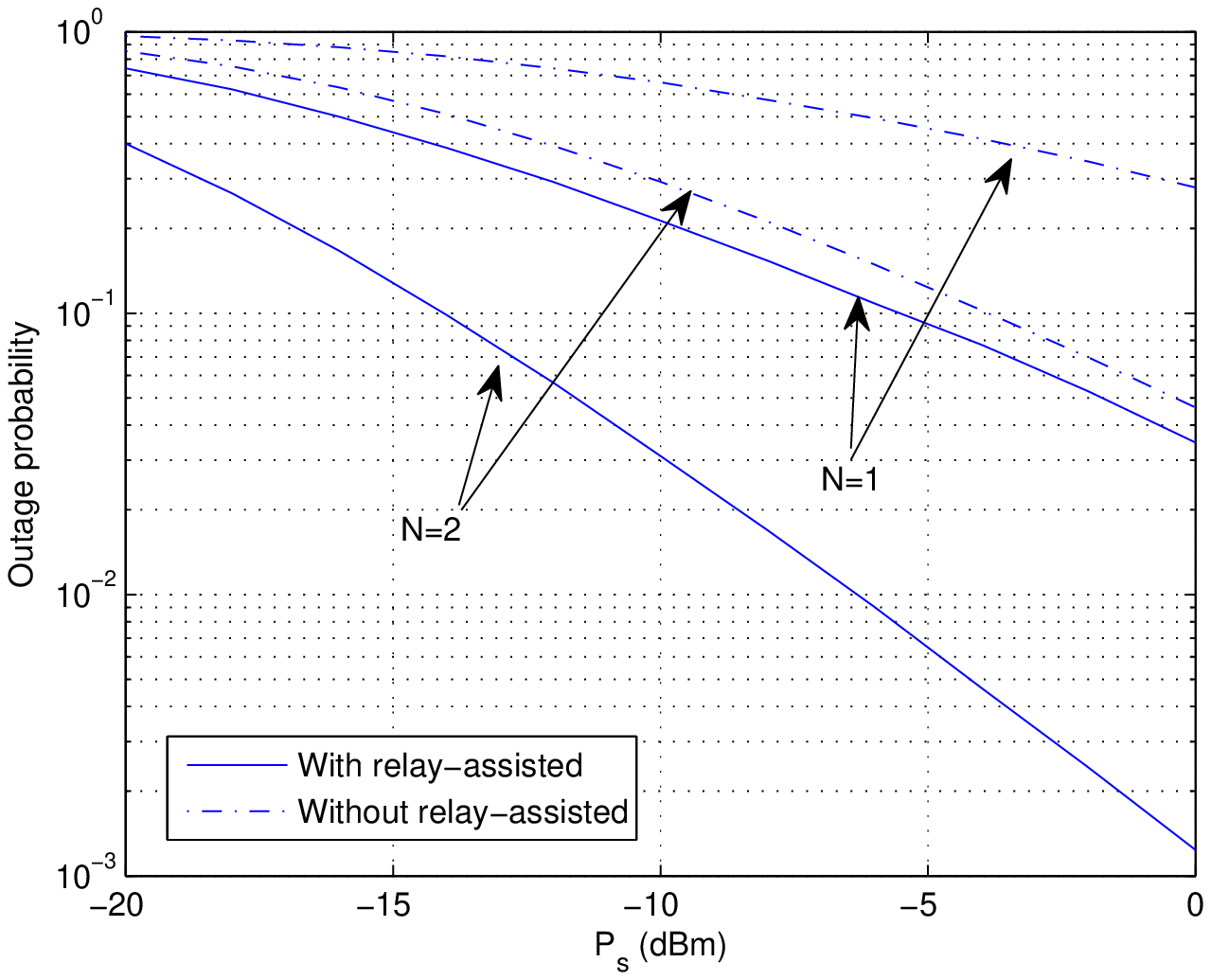}}
  \caption{Impact of relay link on the system performance.}\label{fig:Impact of relay0}
\end{figure}

Fig. \ref{fig:Impact of relay0} compares the achievable system performance with/without relay link when $d_1=30$ m, $d_2=d_3=16$ m $\alpha=3$, $\tau=0.5$. As shown in Fig. \ref{fig:01}, in terms of throughput, using the relay link maybe beneficial in the low SNR regime, while becomes worse as the SNR increases. This is because the use of relay transmission occupies an extra time slot, which incurs a loss of spectral efficiency in the high SNR regime. Also, as shown in Fig. \ref{fig:02}, in terms of outage probability, using the relay link always leads to superior outage performance.when it comes to the throughput performance. This is intuitive since the use of relay transmission provide extra cooperative diversity, which improves the link reliability.

\begin{figure}[!ht]
  \centering
  \subfigure[Outage probability]{\label{fig:outage}\includegraphics[scale=0.5]{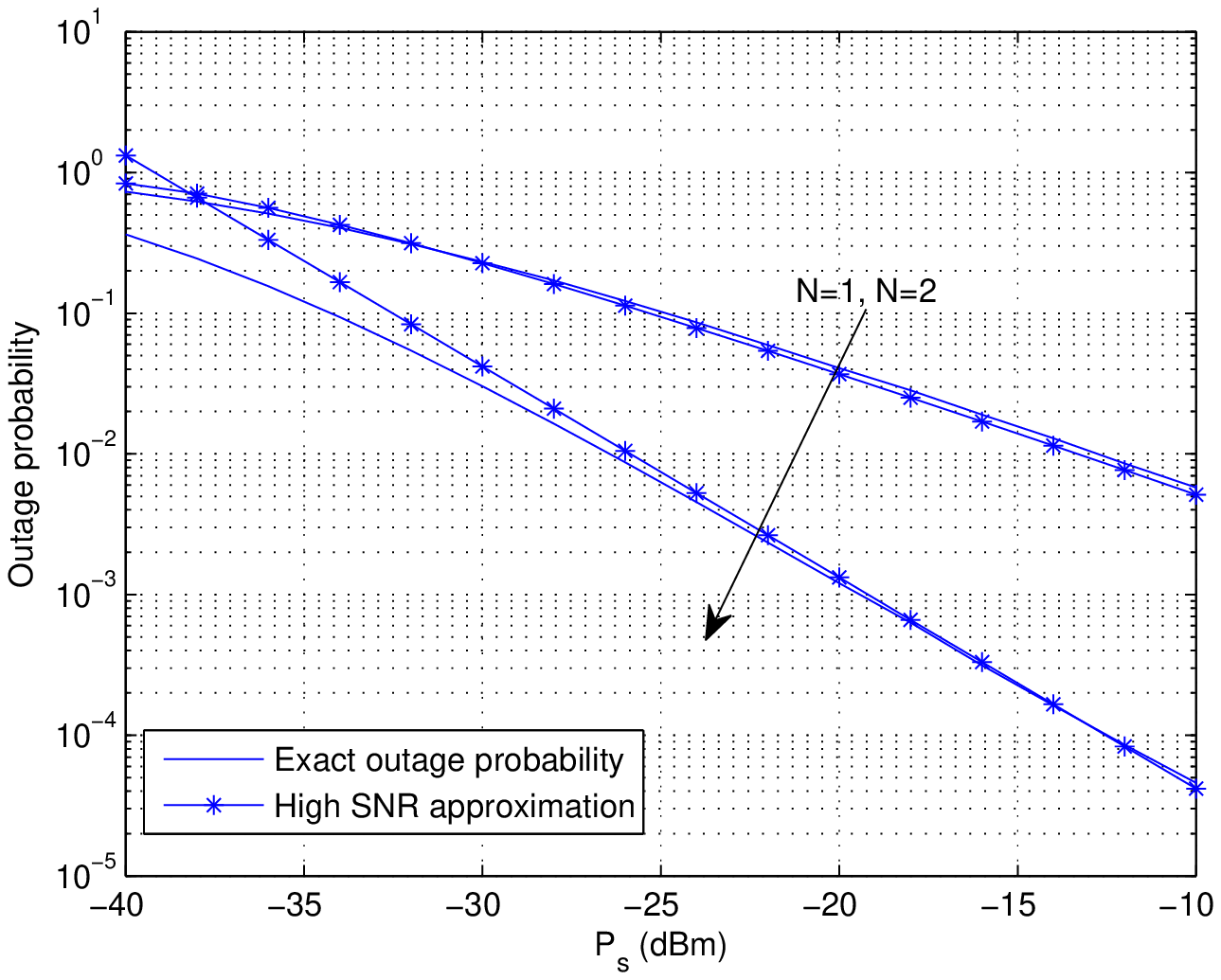}}
  \subfigure[Average throughput]{\label{fig:throughput_monte_vs_lower}\includegraphics[scale=0.5]{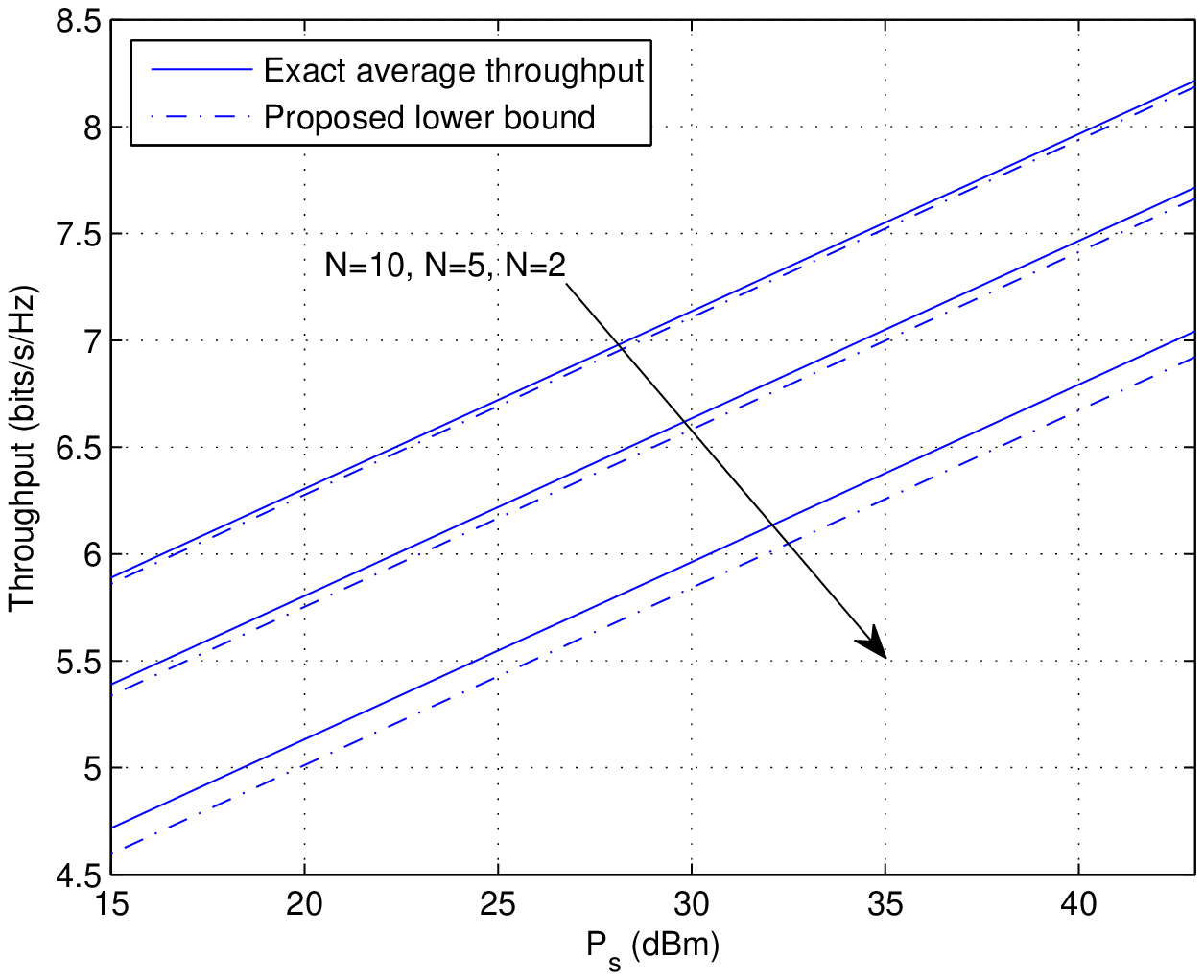}}
  \caption{Impact of $N$ on the system performance.}\label{fig:system performance}
\end{figure}

{Fig. \ref{fig:outage} and Fig. \ref{fig:throughput_monte_vs_lower} show the impact of $N$ on the outage performance and average throughput of the system with $d_1=20$ m, $d_2=d_3=15$ m and the given outage threshold is $0$ dB. It can be readily observed from Fig. \ref{fig:outage} that increasing the number of antennas significantly improves the outage performance. Also, the high SNR approximations in (\ref{eq:highsnr}) are quite accurate. In addition, it is observed that the system achieves a diversity order of $\frac{N+1}{2}$ which is consistent with our analytical results derived in Theorem \ref{theorem:highsnr}. Finally, Fig. \ref{fig:throughput_monte_vs_lower} indicates that increasing $N$ can substantially boost the average throughput, and the proposed analytical lower bound in (\ref{lower bound multi}) is sufficiently tight across the entire SNR region of interest, especially when $N$ is large.}

\section{conclusion}\label{section:conclusion}

We have presented a detailed investigation on the performance of a three-node wireless powered relay system. Specifically, we studied the optimal beamformer and time split design. To circumvent the computation complexity of the original optimization problem, we proposed a suboptimal beamformer and time split solution by optimizing the throughput upper bound, where closed-form expressions were obtained. In addition, we looked into the asymptotic large antenna regime, and derived closed-form expressions for the optimal beamformer and time split. Finally, to analytically characterise the achievable system performance, we considered an intuitive beamforming scheme by directing all the energy toward the user, and provided an in-depth study of the outage probability and average throughput. The outcome of the paper indicates that implementing multiple antennas at the H-AP can significantly improve the system performance, and the near optimal performance can be attained by employing the simple suboptimal energy beamforming vector and time split.


\appendices
\section{Proof of Theorem \ref{theorem outmulti}}\label{appendix:theorem outmulti}
The end-to-end SNR in (\ref{total SNR multi}) can be alternatively expressed as
\begin{align}\label{total SNR multi2}
\gamma_{\sf m}=a_1\left\|{\bf{h}}_1\right\|^4
+\frac{{b_1c_1\left\|{\bf{h}}_1\right\|^2\left|h_3\right|^2\nu\left\|{\bf{h}}_2\right\|^4}}{{b_1\left\|{\bf{h}}_1\right\|^2\left|h_3\right|^2}+{c_1\nu\left\|{\bf{h}}_2\right\|^4}+1},
\end{align}
where $\nu=\frac{\left|{\bf{h}}_2^T {\bf{h}}_1^{*}\right|^2}{\left\|{\bf{h}}_2\right\|^2\left\|{\bf{h}}_1\right\|^2}$.

According to \cite{C.K.Au-Yeung}, the random variable $\nu\in\left[0,1\right]$ is independent with ${\bf{h}}_1$ and ${\bf{h}}_2$ with probability density function
\begin{align}\label{pdfofnu}
f(\nu)=\left(N-1\right)\left(1-\nu\right)^{N-2}.
\end{align}
It is easy to observe that $\left\|{\bf{h}}_1\right\|^2$ and $\left\|{\bf{h}}_2\right\|^2$ are i.i.d. chi-square random variables and $\left|h_3\right|^2$ is an exponential random variable, hence, with the help of \cite[Eq. (8.352.4)]{Tables} and \cite[Eq. (8.350.2)]{Tables}, the CDF of $\nu\left\|{\bf{h}}_2\right\|^4$ can be obtained as
\begin{align}\label{nuh2}
F_{\nu\left\|{\bf{h}}_2\right\|^4}(x)=1-2\sum_{m=0}^{N-1}\sum_{i=0}^{N-2}\binom{N-2}{i}\frac{(-1)^i\left(N-1\right)}{m!}x^{i+1}\Gamma\left(m-2i-2,\sqrt{x}\right).
\end{align}

Then, defining $y\triangleq \left\|{\bf{h}}_1\right\|^2$, $\mu\triangleq \left|h_3\right|^2$ and $z \triangleq \nu\left\|{\bf{h}}_2\right\|^4$, substituting (\ref{total SNR multi2}) into (\ref{outage definition}), after some simple algebraic manipulations, the outage probability can be expressed as
\begin{align}
P_{\sf out}=P\left(zk_1<k_2\right),
\end{align}
where $k_1=b_1c_1y\mu +a_1c_1y^2-\gamma_{\sf th}c_1$ and $k_2=\left(\gamma_{\sf th}-a_1y^2\right)\left(b_1y\mu +1\right)$.

Now, depending on the sign of $k_1$ nd $k_2$, the outage probability can be computed by considering four separate cases. Since $z$ is positive, $P\left(zk_1<k_2\right)=0$ when $k_1>0, k_2<0$. When $k_1<0, k_2<0$, we have $\mu<\frac{\gamma_{\sf th}-a_1y^2}{b_1y}$ and $\gamma_{\sf th}-a_1y^2<0$. Again, this is not possible since $\mu$ is non-negative. Therefore, we only need to consider the remaining two cases $k_1<0, k_2>0$ and $k_1>0, k_2>0$. The desired results can then be obtained after some simple algebraic manipulations.

\section{Proof of Theorem \ref{theorem:highsnr}}\label{appendix:theorem:highsnr}
Since the inequality in (\ref{upper bound}) becomes almost exact in the high SNR regime, the outage probability in (\ref{outage definition}) can be approximated as
\begin{align}
P_{\sf out}\approx P_{\sf out}^{\infty}=P\left(a_1\left\|{\bf{h}}_1\right\|^4+\min\left\{b_1\left\|{\bf{h}}_1\right\|^2\left|h_3\right|, c_1\nu\left\|{\bf{h}}_2\right\|^4\right\}<\gamma_{\sf th}\right),
\end{align}

which can be rewritten as
\begin{align}
P_{\sf out}^{\infty}=1-P\left(\min\left\{b_1\left\|{\bf{h}}_1\right\|^2\left|h_3\right|, c_1\nu\left\|{\bf{h}}_2\right\|^4\right\}>\gamma_{\sf th}-a_1\left\|{\bf{h}}_1\right\|^4\right),
\end{align}

Conditioned on $\left\|{\bf{h}}_1\right\|^2=y$, we have
\begin{multline}\label{pout highSNR}
P_{\sf out}^{\infty}=1-\mathbb{E} \left[P\left(\left|h_3\right|^2>\frac{\gamma_{\sf th}-a_1y^2}{b_1y}\right)P\left(v\left\|{\bf{h}}_2\right\|^4>\frac{\gamma_{\sf th}-a_1y^2}{c_1}\right)\bigg|\left\|{\bf{h}}_1\right\|^2=y, \gamma_{\sf th}-ay^2>0\right]\\
-P\left(\gamma_{\sf th}-a\left\|{\bf{h}}_1\right\|^4<0\right).
\end{multline}

With the help of (\ref{nuh2}), (\ref{pout highSNR}) can be expressed as
\begin{multline}\label{outage multi}
P_{\sf out}^{\infty}=1-2\sum_{m=0}^{N-1}\sum_{i=0}^{N-2}\binom{N-2}{i}\frac{(-1)^i\left(N-1\right)}{m!}\int_0^{\sqrt{\frac{\gamma_{\sf th}}{a_1}}}e^{-\frac{\gamma_{\sf th}}{b_1y}}\left(\frac{\gamma_{\sf th}-a_1y^2}{c_1}\right)^{i+1}\\
\times\Gamma\left(m-2i-2,\sqrt{\frac{\gamma_{\sf th}-a_1y^2}{c_1}}\right)
\frac{y^{N-1}}{\Gamma(N)}e^{\frac{a_1-b_1}{b_1}y}dy
-\frac{\Gamma\left(N,\sqrt{\frac{\gamma_{\sf th}}{a_1}}\right)}{\Gamma(N)}.
\end{multline}
Note that the expansion of $\Gamma\left(m-2i-2,\sqrt{\frac{\gamma_{\sf th}-a_1y^2}{c_1}}\right)$ depends on the sign of $m-2i-2$. Hence, it is convenient to consider two separate cases, i.e., $m-2i-2\leqslant0$ and $m-2i-2>0$\footnote{Due to the range of $m$ and $i$, i.e., $m\in[0,N-1]$ and $i\in[0,N-2]$, the case $m-2i-2>0$ occurs if and only if $N\geqslant4$.}.

\subsection*{Case:$m-2i-2\leqslant0$}
When $m-2i-2\leqslant0$, utilizing \cite[Eq. (8.4.15)]{Handbook}, the integrand in (\ref{outage multi}) can be expressed as in (\ref{I123}), shown at the top of the next page.

\begin{figure*}
\begin{align}
&2\sum_{m=0}^{N-1}\sum_{i=0}^{N-2}\binom{N-2}{i}\frac{(-1)^i\left(N-1\right)}{m!}\int_0^{\sqrt{\frac{\gamma_{\sf th}}{a_1}}}e^{-\frac{\gamma_{\sf th}}{b_1y}}\left(\frac{\gamma_{\sf th}-a_1y^2}{c_1}\right)^{i+1}\\
&~~~~~~~~~~~~~~~~~~~~~~~~~~~~~~~~~~~~~~~~~~~~~~\times\Gamma\left(m-2i-2,\sqrt{\frac{\gamma_{\sf th}-a_1y^2}{c_1}}\right)\frac{y^{N-1}}{\Gamma(N)}e^{\frac{a_1-b_1}{b_1}y}dy\nonumber\\
&=2\sum_{m=0}^{N-1}\sum_{i=0}^{N-2}\binom{N-2}{i}\frac{(-1)^{2+3i-m}\left(N-1\right)}{m!\left(2+2i-m\right)!}\psi(3+2i-m)\nonumber\\
&~~~\underbrace{~~~~~~~~~~~~~~~~~~~~~~~~~~~~~~~~~~~~~~~~~~~~~\times\int_0^{\sqrt{\frac{\gamma_{\sf th}}{a_1}}}e^{-\frac{\gamma_{\sf th}}{b_1y}}\left(\frac{\gamma_{\sf th}-a_1y^2}{c_1}\right)^{i+1}\frac{y^{N-1}}{\Gamma(N)}e^{\frac{a_1-b_1}{b_1}y}dy\nonumber}_{I_1}\\
&-\sum_{m=0}^{N-1}\sum_{i=0}^{N-2}\binom{N-2}{i}\frac{(-1)^{2+3i-m}\left(N-1\right)}{m!\left(2+2i-m\right)!}\nonumber\\
&~~~\underbrace{~~~~~~~~~~~~~~~~~~~~~~~~~~~~\times\int_0^{\sqrt{\frac{\gamma_{\sf th}}{a_1}}}e^{-\frac{\gamma_{\sf th}}{b_1y}}\left(\frac{\gamma_{\sf th}-a_1y^2}{c_1}\right)^{i+1}\ln\frac{\gamma_{\sf th}-a_1y^2}{c_1}\frac{y^{N-1}}{\Gamma(N)}e^{\frac{a_1-b_1}{b_1}y}dy\nonumber}_{I_2}\\
&-2\sum_{m=0}^{N-1}\sum_{i=0}^{N-2}\sum_{\begin{subarray}{c}k=0\\k\ne -m+2i+2\end{subarray}}^{\infty}\binom{N-2}{i}\frac{(-1)^{i+k}\left(N-1\right)}{m!k!(k+m-2i-2)}\nonumber\\
&~~~\underbrace{~~~~~~~~~~~~~~~~~~~~~~~~~~~~~~~~~~~~~~~~~~\times\int_0^{\sqrt{\frac{\gamma_{\sf th}}{a_1}}}e^{-\frac{\gamma_{\sf th}}{b_1y}}\left(\frac{\gamma_{\sf th}-a_1y^2}{c_1}\right)^{\frac{m+k}{2}}\frac{y^{N-1}}{\Gamma(N)}e^{\frac{a_1-b_1}{b_1}y}dy\nonumber}_{I_3}\\
&=I_1-I_2-I_3.\label{I123}
\end{align}
\hrule
\end{figure*}

We start with $I_3$. Noticing that $\frac{\gamma_{\sf th}-a_1y^2}{c_1}\rightarrow0$ as $\rho\rightarrow\infty$, it suffices to consider the most significant terms, i.e., $m=0,k=0$, $m=0,k=1$ and $m=1,k=0$. Further noticing that the sum of the two terms $m=0,k=1$ and $m=1,k=0$ is zero due to opposite sign, it remains to focus on the term $m=0, k=0$. Utilizing the Taylor expansion on $e^{\frac{a_1-b_1}{b_1}y}$, $I_3$ can be approximated as
\begin{align}
I_3&\approx
2\sum_{i=0}^{N-2}\binom{N-2}{i}\frac{(-1)^i\left(N-1\right)}{\Gamma(N)(2i+2)}\sum_{l_1=0}^{\infty}\left(\frac{a_1-b_1}{b_1}\right)^{l_1}\frac{1}{l_1!}\left(\frac{\gamma_{\sf th}}{b_1}\right)^{N+l_1}\sum_{k=0}^{\infty}\frac{\left(-\frac{\sqrt{\gamma_{\sf th}a_1}}{b_1}\right)^k}{k!\left(k-(N+l_1)\right)}\nonumber\\
&-2\sum_{i=0}^{N-2}\binom{N-2}{i}\frac{(-1)^i\left(N-1\right)}{\Gamma(N)(2i+2)}\sum_{l_2=0}^{\infty}\left(\frac{a_1-b_1}{b_1}\right)^{l_2}\frac{1}{l_2!}\nonumber\\
&~~~~~~~~~~~~~~~~~~~~~~~~~~~~~~~~\times\left(\frac{\gamma_{\sf th}}{b_1}\right)^{N+l_2}\frac{(-1)^{N+l_2}}{(N+1)!}\left(\psi(N+l_2+1)-\ln\frac{\sqrt{\gamma_{\sf th}a_1}}{b_1}\right).\label{eq22}
\end{align}

Noticing that $2\sum_{i=0}^{N-2}\binom{N-2}{i}\frac{(-1)^i\left(N-1\right)}{\Gamma(N)(2i+2)}=1$, it is easy to show that the most significant terms in (\ref{eq22}) appears in the first part of (\ref{eq22}) when $l_1=0$ and $k=0$. Hence, omitting the higher order terms of $\frac{1}{\rho}$, (\ref{eq22}) can be approximated as
\begin{align}
I_3&\approx-\frac{1}{N!}\left(\frac{\gamma_{\sf th}}{a_1}\right)^{\frac{N}{2}}+\frac{1}{(N-1)!}\left(\frac{\gamma_{\sf th}}{a_1}\right)^{\frac{N+1}{2}}\frac{2a_1+b_1(N-1)}{b_1(N+1)(N-1)}.
\end{align}

To this end, using a similar approach, it can be shown that the most significant terms in $I_1$ and $I_2$ is on the order of $\frac{1}{\rho}^{\frac{N+2}{2}}$, which implies that $I_3$ is the dominant term.

\subsection*{Case:$m-2i-2>0$}
When $m-2i-2>0$, expanding $\Gamma\left(m-2i-2,\sqrt{\frac{\gamma_{\sf th}-a_1y^2}{c_1}}\right)$ with the help of \cite[Eq. (8.354.2)]{Tables}, and following the similar approach as in the case $m-2i-2\leqslant0$, it can be shown that the minimum order of $\frac{1}{\rho}$ is $\frac{1}{\rho}^{\frac{N+2}{2}}$. Therefore, all the terms from $m-2i-2>0$ are insignificant compared to $I_3$. Therefore, $P_{\sf out}^{\infty}$ can be approximated by
\begin{multline}\label{outage multi111}
P_{\sf out}^{\infty}\approx 1-\frac{1}{N!}\left(\frac{\gamma_{\sf th}}{a_1}\right)^{\frac{N}{2}}+\frac{1}{(N-1)!}\left(\frac{\gamma_{\sf th}}{a_1}\right)^{\frac{N+1}{2}}\frac{2a_1+b_1(N-1)}{b_1(N+1)(N-1)}
-\frac{\Gamma\left(N,\sqrt{\frac{\gamma_{\sf th}}{a_1}}\right)}{\Gamma(N)}.
\end{multline}
Then, using \cite[Eq. (8.354.2)]{Tables}, $\frac{\Gamma\left(N,\sqrt{\frac{\gamma_{\sf th}}{a_1}}\right)}{\Gamma(N)}$ can be approximated by
\begin{align}
\frac{\Gamma\left(N,\sqrt{\frac{\gamma_{\sf th}}{a_1}}\right)}{\Gamma(N)}\approx1-\frac{1}{N!}\left(\frac{\gamma_{\sf th}}{a_1}\right)^\frac{N}{2}+\frac{1}{(N-1)!(N+1)}\left(\frac{\gamma_{\sf th}}{a_1}\right)^\frac{N+1}{2}\label{Ga}.
\end{align}
To this end, substituting (\ref{Ga}) into (\ref{outage multi111}), the desired result can be obtained after some simple algebraic manipulations.

\section{Proof of Theorem \ref{theorem:throughput}}\label{appendix:theorem:throughput}
Using (\ref{convex approximation}), the average throughput can be lower bounded by
\begin{align}\label{ergodic upperbound}
C\geqslant\frac{1-\tau}{2}\log_2\left(1+e^{\mathbb{E}\left[\ln\gamma_{us}\right]}+e^{\mathbb{E}\left[\ln\gamma_{ur}\right]+\mathbb{E}\left[\ln\gamma_{rs}\right]-\mathbb{E}\left[\ln\left(1+\gamma_{ur}+\gamma_{rs}\right)\right]}\right).
\end{align}
Then, applying the Jensen's inequality on the term $\mathbb{E}\left[\ln\left(1+\gamma_{ur}+\gamma_{rs}\right)\right]$, we have the following lower bound
\begin{align}\label{ergodic approximation}
C_{\sf low}=\frac{1-\tau}{2}\log_2\left(1+e^{\mathbb{E}\left[\ln\gamma_{us}\right]}+
e^{\mathbb{E}\left[\ln\gamma_{ur}\right]+\mathbb{E}\left[\ln\gamma_{rs}\right]-\ln\left(1+\mathbb{E}\left[\gamma_{ur}\right]
+\mathbb{E}\left[\gamma_{rs}\right]\right)}\right).
\end{align}
Hence, the remaining task is to compute $\mathbb{E}\left[\ln\gamma_{us}\right]$, $\mathbb{E}\left[\ln\gamma_{ur}\right]$, $\mathbb{E}\left[\ln\gamma_{rs}\right]$, $\mathbb{E}\left[\gamma_{ur}\right]$ and $\mathbb{E}\left[\gamma_{rs}\right]$.

Noticing that the expectation of $\ln\gamma_{t}$, $t\in \left\{us, ur, rs\right\}$ can be derived from
\begin{align}\label{expectation ln}
\mathbb{E}\left[\ln\gamma_{t}\right]=\frac{d\mathbb{E}\left[\gamma_{t}^n\right]}{dn}\bigg{|}_{n=0},
\end{align}
while the general moment can be obtained by
\begin{align}\label{moment of gamma}
\mathbb{E}\left[\gamma_{t}^n\right]=\int_0^\infty x^n f_{\gamma_t}(x)dx = n\int_0^\infty x^{n-1}\left(1-F_{\gamma_t}(x)\right)dx.
\end{align}
Hence, the key task is to obtain closed-form expressions for the CDF of ${\gamma_t}$, $t\in \left\{us, ur, rs\right\}$, which we do in the following.

It is easy to show that the CDF of ${\gamma_{us}}$ is given by
\begin{align}\label{cdf gamma_us}
F_{\gamma_{us}}(x)=P\left(\left\|{\bf{h}}_1\right\|^2<\sqrt{\frac{x}{a_1}}\right)=1-\frac{\Gamma\left(N,\sqrt{\frac{x}{a_1}}\right)}{\Gamma(N)}.
\end{align}
Also, with the help of (\ref{nuh2}), the CDF of ${\gamma_{rs}}$ can be obtained as
\begin{align}\label{cdf gamma_rs}
F_{\gamma_{rs}}(x)=1-2\left(N-1\right)\sum_{m=0}^{N-1}\frac{1}{m!}\sum_{i=0}^{N-2}\binom{N-2}{i}(-1)^i \left(\frac{x}{c_1}\right)^{i+1}\Gamma\left(m-2i-2,\sqrt{\frac{x}{\nu}}\right).
\end{align}
Similarly, the help of \cite[Eq. (8.352.4)]{Tables} and \cite[Eq. (8.432.7)]{Tables}, the CDF of ${\gamma_{ur}}$ can be derived as
\begin{align}
F_{\gamma_{ur}}(x)
&=1-\sum_{m=0}^{N-1}\frac{2}{m!}\left({\frac{x}{b_1}}\right)^{\frac{m+1}{2}}K_{m-1}\left(2\sqrt{\frac{x}{b_1}}\right)\label{eq32},
\end{align}

Then, substituting (\ref{cdf gamma_us}), (\ref{cdf gamma_rs}) and (\ref{eq32}) into (\ref{moment of gamma}), the general moment of ${\gamma_t}$, $t\in \left\{us, ur, rs\right\}$ can be computed as shown in the following. To start, the general moment of $\gamma_{us}^n$ can be expressed as
\begin{align}
\mathbb{E}\left[\gamma_{us}^n\right]&=n\int_0^\infty x^{n-1}\frac{\Gamma\left(N,\sqrt{\frac{x}{a_1}}\right)}{\Gamma(N)}dx=2n\sum_{m=0}^{N-1}\frac{1}{m!}a_1^{n}\Gamma(2n+m)\label{eq42},
\end{align}
where we have used \cite[Eq. (8.352.4)]{Tables} and \cite[Eq. (8.310.1)]{Tables}.

And the general moment of $\gamma_{rs}$ can be obtained as
\begin{align}
\mathbb{E}\left[\gamma_{rs}^n\right]&=n2\left(N-1\right)\sum_{m=0}^{N-1}\frac{1}{m!}\sum_{i=0}^{N-2}\binom{N-2}{i}(-1)^i\int_0^\infty x^{n-1}\left(\frac{x}{c_1}\right)^{i+1}\Gamma\left(m-2i-2,\sqrt{\frac{x}{\nu}}\right)\label{eq50}\\
&=4n\left(N-1\right)\sum_{m=0}^{N-1}\frac{1}{m!}\sum_{i=0}^{N-2}\binom{N-2}{i}(-1)^i c_1^n\frac{\Gamma(m+2n)}{2n+2i+2}\label{eq51},
\end{align}
where (\ref{eq51}) is obtained by making a change of variable $t=\sqrt{\frac{x}{c_1}}$ with the help of \cite[Eq. (8.14.4)]{Handbook}.

Finally, with the help of \cite[Eq. (9.34.3)]{Tables} and \cite[Eq. (7.811.4)]{Tables}, the general moment of $\gamma_{ur}$ can be computed by
\begin{align}
\mathbb{E}\left[\gamma_{ur}^n\right]
&=nb_1^n\sum_{m=0}^{N-1}\frac{1}{m!}\Gamma(m+n)\Gamma(n+1)\label{eq62}.
\end{align}

To this end, substituting (\ref{eq42}), (\ref{eq51}) and (\ref{eq62}) into (\ref{expectation ln}), the desired result can be obtained after some algebraic manipulations.

\nocite{*}
\bibliographystyle{IEEE}

\begin{thebibliography}{1}

\bibitem{M.Xie}
M. Xia and S. Aissa, ``On the efficiency of far-field wireless power transfer,'' {\em IEEE Trans. Sig. Process.}, vol. 63, no. 11, pp. 2835--2847, Jun. 2015.

\bibitem{S.Bi}
S. Bi, C. K. Ho, and R. Zhang, ``Wireless powered communication: Opportunities and challenges,'' {\em IEEE Commun. Mag.}, vol. 53, no. 4, pp. 117--125, Apr. 2013.

%

\bibitem{R.Zhang0}
R. Zhang and C. Ho, ``MIMO broadcasting for simultaneous wireless information and power transfer,'' {\em IEEE Trans. Wireless Commun.}, vol. 12, no. 5, pp. 1989--2001, May 2013.

%

\bibitem{K.Huang2}
K. Huang and E. G. Larsson, ``Simultaneous information and power transfer for broadband wireless systems,'' {\em IEEE Trans. Signal Process.}, vol. 61, no. 23, pp. 5972--5986, Dec. 2013.

\bibitem{H.Ju0}
H. Ju and R. Zhang, ``Optimal resource allocation in full-duplex wireless-powered communication network,'' {\em IEEE Trans. Commun.}, vol. 62, no. 10, pp. 3528--3540, Oct. 2014.


\bibitem{K.Huang}
K. Huang and V. K. N. Lau, ``Enabling wireless power transfer in cellular networks: Architecture, modeling and deployment,'' {\em IEEE Trans. Wireless Commun.}, vol. 13, no. 2, pp. 902--912, Feb. 2014.
%
%

\bibitem{H.Ju}
H. Ju and R. Zhang, ``Throughput maximization for wireless powered communication networks,'' {\em IEEE Trans. Wireless Commun.}, vol. 13, no. 1, pp. 418--428, Jan. 2014.

\bibitem{Y.L.Che}
Y. L. Che, L. Duan, and R. Zhang, ``Spatial throughput maximization of wireless powered communication networks,'' {\em  IEEE J. Sel. Areas Commum.}, vol. 33, no. 8, pp. 1534--1548, Aug. 2015.

\bibitem{Q.Wu}
Q. Wu, M. Tao, D. W. K. Ng, W. Chen, and R. Schober, ``Energy-efficient resource allocation for wireless powered communication,'' {\em IEEE Trans. Wireless Commun.}, vol. 15, no. 3, pp. 2312--2327, Mar. 2016.

\bibitem{L.Liu2}
L. Liu, R. Zhang and K. Chua, ``Multi-antenna wireless powered communication with energy beamforming,'' {\em IEEE Trans. Commun.}, vol. 62, no. 12, pp. 4349--4361, Dec. 2014.

\bibitem{X.Chen0}
X. Chen, C. Yuen, and Z. Zhang, ``Wireless energy and information transfer tradeoff for limited feedback multi-antenna systems with energy beamforming,'' {\em IEEE Trans. Veh. Tech.}, vol. 63, no. 1, pp. 407--412, Jan. 2014.

\bibitem{W.Huang}
W. Huang, H. Chen, Y. Li, and B. Vucetic, ``On the performance of multi-antenna wireless-powered communications with energy beamforming,'' {\em IEEE Trans. Veh. Techno.}, vol. 65, no. 3, pp. 1801-1808, Mar. 2016.

\bibitem{C.Zhong}
C. Zhong, X. Chen, Z. Zhang, and G. K. Karagiannidis, ``Wireless-powered communications: Performance analysis and optimization,'' {\em IEEE Trans. Commun.}, vol. 63, no. 12, pp. 5178-5190, Dec. 2015.

\bibitem{C.Zhong0}
C. Zhong, G. Zheng, Z. Zhang, and G. Karagiannids, ``Optimum wirelessly powered relaying,'' {\em IEEE Signal Process. Lett.}, vol. 22, no. 10, pp. 1728--1732, Oct. 2015.

\bibitem{H.Ju3}
H. Ju and R. Zhang, ``User cooperation in wireless powered communication networks,'' in {\em Proc. IEEE GLOBECOM}, Dec. 2014, pp. 1430--1435.

\bibitem{Z.Ding}
Z. Ding, C. Zhong, D. W. K. Ng, M. Peng, H. A. Suraweera, R. Schober, and H. V. Poor, ``Application of smart antenna technologies in simultaneous wireless information and power transfer,''  {\em IEEE Commun. Mag.}, vol, 53, no. 4, pp. 86--93, Apr. 2015.

\bibitem{H.Chen}
H. Chen, Y. Li, J. L. Rebelatto, B. F. Uchoa-Filho, and B. Vucetic, ``Harvest-then-cooperate: Wireless-powered cooperative communications,'' {\em IEEE Trans. Signal Process.}, vol. 63, no. 7, pp. 1700--1711, Apr. 2015.


\bibitem{Y.Huang1}
Y. Huang, and B. Clerckx, ``Relaying strategies for wireless-powered MIMO relay networks,'' {\em IEEE Trans. Wireless Commun.}, vol. 15, no. 9, pp. 6033-6047, Sep. 2016.

\bibitem{SigTel}
H. Liang, C. Zhong, H. A. Suraweera, G. Zheng, and Z. Zhang, ``Beamformer and time split design for wireless powered multi-antenna cooperative systems,'' accepted to appear in SigTelCom 2017.

\bibitem{Tables}
I. S. Gradshteyn and I. M. Ryzhik, {\em Tables of Intergrals, Series and Products,} $6$th Ed.. San Diego: Academic Press, 2000.

\bibitem{Y.Zeng2}
Y. Zeng and R. Zhang, ``Full-duplex wireless-powered relay with self-energy recycling,'' {\em IEEE Wireless Commun. Lett.}, vol. 4, no. 2, pp. 201--204, Apr. 2015.

\bibitem{Y.Zeng}
Y. Zeng and R. Zhang, ``Optimized training design for wireless energy transfer,'' {\em IEEE Trans. Commun.}, vol. 63, no. 2, pp. 536--550, Feb. 2015.

\bibitem{F.Gao}
F. Gao, T. Cui, and A. Nallanathan, ``On channel estimation and optimal training design for amplify and forward relay network,'' {\em IEEE Trans. Wireless Commun.}, vol. 7,  no. 5, pp. 1907--1916, May 2008.

%

\bibitem{A.Nasir}
A. A. Nasir, X. Zhou, S. Durrani, and R. A. Kennedy, ``Relaying protocols for wireless energy harvesting and information processing,'' {\em IEEE Trans. Wireless Commun.}, vol. 12, no. 7, pp. 3622--3636, July. 2013.


\bibitem{E.A.Jorswieck}
E. A. Jorswieck, E. G. Larsson, and D. Danev, ``Complete characterization of the pareto boundary for the MISO interference channel,'' {\em IEEE Trans. Signal Process.}, vol. 56, no. 10, pp. 5292--5296, Oct. 2008.

\bibitem{S.Ikki}
S. Ikki and M. H. Ahmed, ``Performance analysis of cooperative diversity wireless networks over nakagami-m fading channel,'' {\em IEEE Commum. Lett.}, vol. 11, no. 4, pp. 334--336, Apr. 2007.

\bibitem{E.A.Jorswieck2}
E. A. Jorswieck and E. G. Larsson, ``The MISO interference channel from a game-theoretic perspective: A combination of selfishness and altruism achieves Pareto optimality,'' in {\em Proc. Int. Conf. Acoustics, Speech, Signal Processing (ICASSP)}, 2008.

\bibitem{G.L.Moritz}
G. L. Moritz, J. L. Rebelatto, R. D. Souza, B. F. Uchoa-Filho and Y. Li, ``Time-Switching Uplink Network-Coded Cooperative Communication With Downlink Energy Transfer,'' {\em IEEE Trans. Signal Process.}, vol. 62, no. 19, pp. 5009--5019, Oct. 2014.

\bibitem{C.Zhong02}
C. Zhong, H. A. Suraweera, G. Zheng, I. Krikidis, and Z. Zhang, ``Wireless information and power transfer with full duplex relaying,'' {\em IEEE Trans. Commun.}, vol. 62, no. 10, pp. 3447--3461, Oct. 2014.

\bibitem{T.L.Marzetta}
T. L. Marzetta, ``Noncooperative cellular wireless with unlimited numbers of base station antennas,'' {\em IEEE Trans. Wireless Commun.}, vol. 9, no. 11, pp. 3590--3600, Nov. 2010.


\bibitem{H.Xie2}
H. Xie, F. Gao, and S. Jin, ``An overview of low-rank channel estimation for massive MIMO systems,'' {\em IEEE Access}, vol. 4, pp. 7313--7321, Nov. 2016.

\bibitem{H.Xie3}
H. Xie, B. Wang, F. Gao, and S. Jin, ``A full-space spectrum-sharing strategy for massive MIMO cognitive radio,'' {\em IEEE J. Select. Areas Commun.}, vol. 34, no. 10, pp. 2537--2549, Oct. 2016.

\bibitem{X.Chen1}
X. Chen, X. Wang, and X. Chen, ``Energy-efficient optimization for wireless information and power transfer in large-scale MIMO systems employing energy beamforming,'' {\em IEEE Wireless Commun. Lett.}, vol. 2, no. 6, pp. 667--670, Dec. 2013.


{\bibitem{D.W.K.Ng}
D. W. K. Ng, E. S. Lo, and R. Schober, ``Wireless information and power transfer: energy efficiency optimization in OFDMA systems,'' {\em IEEE Trans. Wireless Commun.}, vol. 12, no. 12, pp. 6352--6370, Dec. 2013.}

\bibitem{C.K.Au-Yeung}
C. K. Au-Yeung and D. J. Love, ``On the performance of random cector quantization limited feedback beamforming in a MISO system,'' {\em IEEE Trans. Wireless Commun.}, vol. 6, no. 2, pp. 458--462, Feb. 2007.

\bibitem{Handbook}
F. W. J. Olver, D. W. Lozier, R. F. Boisvert and C. W. Clark, {\em Handbook of Mathematical Functions,} Cambridge University Press, 2010.

\end{thebibliography}
\begin{footnotesize}

\end{footnotesize}

\end{document}